\documentclass[10pt,twocolumn]{article}
\usepackage[top=1.9cm, bottom=1.9cm, left=1.7cm, right=1.7cm]{geometry}
\usepackage{times}
\usepackage[hyphens]{url}  %
\usepackage[keeplastbox]{flushend}  %
\usepackage{graphicx}  %
\usepackage{ifthen}
\newif\ifshort
\shortfalse %
\ifshort
  \newcommand{\isShort}{true}
\else
  \newcommand{\isShort}{false}
\fi
\newcommand{\shortVer}[1]{\ifthenelse{\equal{\isShort}{true}}{{#1}}{}}
\newcommand{\longVer}[1]{\ifthenelse{\equal{\isShort}{false}}{{#1}}{}}

\usepackage{xcolor}
\usepackage{flushend}
\usepackage{booktabs}
\usepackage{graphicx}
\usepackage{paralist}
\usepackage{enumitem}
\usepackage[small,bf]{caption}
\usepackage{subfigure}
\usepackage[hang,flushmargin]{footmisc}
\renewcommand{\footnotesize}{\fontsize{8}{9}\selectfont}

\usepackage[bookmarks=true,%
            bookmarksnumbered=true,%
            colorlinks=true,%
            linkcolor=linkcol,%
            citecolor=citecol,%
            urlcolor=urlcol,%
            hypertexnames=true,%
            pdfpagelabels,
	    ]%
      {hyperref}

\usepackage[compact]{titlesec}
\titlespacing*{\section}{0pt}{*4}{4pt} 
\titlespacing{\subsection}{0pt}{*3}{3pt}

\definecolor{linkcol}{rgb}{0,0,0.5}
\definecolor{citecol}{rgb}{0,0.5,0.3}
\definecolor{urlcol}{rgb}{0.3,0,0}

\renewenvironment{thebibliography}[1]{
  \begin{oldthebibliography}{#1}
    \setlength{\itemsep}{0.1em}
    \setlength{\parskip}{0.1em}
}
{
  \end{oldthebibliography}
}

\renewcommand{\footnoterule}{%
  \kern -3pt
  \hrule width 1in
  \kern 2pt
}

\usepackage{xspace}

\usepackage{todonotes} %

\newif\ifcomment
\commenttrue
\ifcomment
\newcommand{\XXX}[2]{{\bf \textcolor{blue}{#1: #2}}}
\newcommand{\jbnote}[1]{{\bf \textcolor{magenta}{JB: #1}}}
\newcommand{\msnote}[1]{{\bf \textcolor{magenta}{MS: #1}}}
\newcommand{\edc}[1]{{\bf \textcolor{blue}{EDC: #1}}}
\newcommand{\gs}[1]{{\bf \textcolor{red}{gs: #1}}}
\newcommand{\tc}[1]{{\bf \textcolor{orange}{TC: #1}}}
\else
\newcommand{\XXX}[2]{}
\newcommand{\jbnote}[1]{}
\newcommand{\msnote}[1]{}
\newcommand{\edc}[1]{}
\newcommand{\gs}[1]{}
\newcommand{\tc}[1]{}
\fi

\newcommand{\descr}[1]{\smallskip\noindent\textbf{#1}}

\newcommand{\dspol}{{/pol/}\xspace}

\makeatletter
\def\url@leostyle{%
  \@ifundefined{selectfont}{\def\UrlFont{}}%
  {\def\UrlFont{}}%
}
\makeatother
\usepackage[hyphenbreaks]{breakurl}

\newif\ifwatermark
\watermarkfalse

\ifwatermark
    \usepackage{draftwatermark}
    \SetWatermarkScale{.7}
    \SetWatermarkText{\shortstack{In Submission:\\Do Not Distribute}}
    \SetWatermarkColor[rgb]{1,0.88,0.88}
\fi

\usepackage{etoolbox}
\makeatletter
\patchcmd\@combinedblfloats{\box\@outputbox}{\unvbox\@outputbox}{}{%
   \errmessage{\noexpand\@combinedblfloats could not be patched}%
}%
 \makeatother
\AtBeginShipout{%
  \ifnum\value{page}>1 %
    \typeout{* Additional boxing of page `\thepage'}%
    \setbox\AtBeginShipoutBox=\hbox{\copy\AtBeginShipoutBox}%
  \fi
}

\begin{document}
\title{\bf Disinformation Warfare: Understanding State-Sponsored Trolls \\ on Twitter and Their Influence on the Web\thanks{A preliminary version of this paper appears in the 4th Workshop on The Fourth Workshop on Computational Methods in Online Misbehavior (CyberSafety 2019) -- WWW'19 Companion Proceedings. This is the full version.}}
\author{Savvas Zannettou$^{1}$, Tristan Caulfield$^2$, Emiliano De Cristofaro$^2$,\\Michael Sirivianos$^{1}$, Gianluca Stringhini$^3$, Jeremy Blackburn$^4$\\[0.5ex]
\normalsize $^1$Cyprus University of Technology, $^2$University College London, $^3$Boston University, $^4$University of Alabama at Birmingham\\
\normalsize sa.zannettou@edu.cut.ac.cy, \{t.caulfield,e.decristofaro\}@ucl.ac.uk,\\[-0.5ex]
\normalsize michael.sirivianos@cut.ac.cy, gian@bu.edu, blackburn@uab.edu}
\date{}

\maketitle

\begin{abstract}
Over the past couple of years, anecdotal evidence has emerged linking coordinated campaigns by state-sponsored actors with efforts to manipulate public opinion on the Web, often around major political events, through dedicated accounts, or ``trolls.''
Although they are often involved in spreading disinformation on social media, there is little understanding of how these trolls operate, what type of content they disseminate, and most importantly their influence on the information ecosystem.

In this paper, we shed light on these questions by analyzing 27K tweets posted by 1K Twitter users identified as having ties with Russia's Internet Research Agency and thus likely state-sponsored trolls.
We compare their behavior to a random set of Twitter users, finding interesting differences in terms of the content they disseminate, the evolution of their account, as well as their general behavior and use of Twitter.
Then, using Hawkes Processes, we quantify the influence that trolls had on the dissemination of news on social platforms like Twitter, Reddit, and 4chan.
Overall, our findings indicate that Russian trolls managed to stay active for long periods of time and to reach a substantial number of Twitter users with their tweets.
When looking at their ability of spreading news content and making it viral, however, we find that their effect on social platforms was minor, with the significant exception of news published by the Russian state-sponsored news outlet RT (Russia Today).
\end{abstract}

\maketitle

\section{Introduction}
Recent political events and elections have been increasingly accompanied by reports of disinformation campaigns attributed to state-sponsored actors~\cite{ferrara2017disinformation}.
In particular, ``troll farms,'' allegedly employed by Russian state agencies, have been actively commenting and posting content on social media to further the Kremlin's political agenda~\cite{independent}.
In late 2017, the US Congress started an investigation on Russian interference in the 2016 US Presidential Election, releasing the IDs of 2.7K Twitter accounts identified as Russian trolls.

Despite the growing relevance of state-sponsored disinformation, the activity of accounts linked to such efforts has not been thoroughly studied.
Previous work has mostly looked at campaigns run by bots~\cite{ferrara2017disinformation,hegelich2016are,ratkiewicz2011detecting};
however, automated content diffusion is only a part of the issue, and in fact recent research has shown that human actors are actually key in spreading false information on Twitter~\cite{starbird2017examining}.
Overall, many aspects of state-sponsored disinformation remain unclear, e.g., how do state-sponsored trolls operate? What kind of content do they disseminate? And, perhaps more importantly, is it possible to quantify the influence they have on the overall information ecosystem on the Web?

In this paper, we aim to address these questions, by relying on the set of 2.7K accounts released by the US Congress as ground truth for Russian state-sponsored trolls.
From a dataset containing all tweets released by the 1\% Twitter Streaming API, we search and retrieve 27K tweets posted by 1K Russian trolls between January 2016 and September 2017.
We characterize their activity by comparing to a random sample of Twitter users.
Then, we quantify the influence of these trolls on the greater Web, looking at occurrences of URLs posted by them on Twitter, 4chan~\cite{hine2016longitudinal}, and Reddit, which we choose since they are impactful actors of the information ecosystem~\cite{zannettou2017web}.
Finally, we use Hawkes Processes~\cite{linderman2014} to model the influence of each Web community (i.e., Russian trolls on Twitter, overall Twitter, Reddit, and 4chan) on each other.

\descr{Main findings.} Our study leads to several key observations:
\begin{enumerate} %
\item Trolls actually bear very small influence in making news go viral on Twitter and other social platforms alike. A noteworthy exception are links to news originating from RT (Russia Today), a state-funded news outlet: indeed, Russian trolls are quite effective in ``pushing'' these URLs on Twitter and other social networks.
\item %
The main topics discussed by Russian trolls target very specific world events (e.g., Charlottesville protests) and organizations (such as ISIS), and political threads related to Donald Trump and Hillary Clinton.
\item Trolls adopt different identities over time, i.e., they ``reset'' their profile by deleting their previous tweets and changing their screen name/information.
\item Trolls exhibit significantly different behaviors compared to other (random) Twitter accounts. For instance, the locations they report concentrate in a few countries like the USA, Germany, and Russia, perhaps in an attempt to appear ``local'' and more effectively manipulate opinions of users from those countries. Also, while random Twitter users mainly tweet from mobile versions of the platform, the majority of the Russian trolls do so via the Web Client.
\end{enumerate}

\longVer{
\section{Background} \label{sec:background}
In this section, we provide a brief overview of the social networks studied in this paper, i.e., Twitter, Reddit, and 4chan,
which we choose because they are impactful actors on the Web's information ecosystem~\cite{zannettou2017web}.

\descr{Twitter.} Twitter is a mainstream social network, where users can broadcast short messages, called ``tweets,'' to their followers.
Tweets may contain hashtags, which enable the easy index and search of messages, as well as mentions, which refer to other users on Twitter.

\descr{Reddit.} Reddit is a news aggregator with several social features.
It allows users to post URLs along with a title; posts can get up-  and down- votes, which dictate the popularity and order in which they appear on the platform.
Reddit is divided to ``subreddits,'' which are forums created by users that focus on a particular topic (e.g., /r/The\_Donald is about discussions around Donald Trump).

\descr{4chan.} 4chan is an imageboard  forum, organized in communities called ``boards,'' each with a different topic of interest.
A user can create a new post by uploading an image with or without some text; others can reply below with or without images.
4chan is an anonymous community, and several of its boards are reportedly responsible for a substantial amount of hateful content~\cite{hine2016longitudinal}.
In this work we focus on the Politically Incorrect board (\dspol) mainly because it is the main board for the discussion of politics and world events.
Furthermore, 4chan is ephemeral, i.e., there is a limited number of active threads and all threads are permanently deleted after a week.
}

\section{Datasets} \label{sec:datasets}

\descr{Russian trolls.} We start from the 2.7K Twitter accounts suspended by Twitter because of connections to Russia's Internet Research Agency. %
The list of these accounts was released by the US Congress as part of their investigation of the alleged Russian interference in the 2016 US presidential election, and includes both Twitter's {\em user ID} (which is a numeric unique identifier associated to the account) and the {\em screen name}.\footnote{See \url{https://democrats-intelligence.house.gov/uploadedfiles/exhibit_b.pdf}}
From a dataset storing all tweets released by the 1\% Twitter Streaming API, we search for tweets posted between January 2016 and September 2017 by the user IDs of the trolls.
Overall, we obtain 27K tweets from 1K out of the 2.7K Russian trolls. %
Note that the criteria used by Twitter to identify these troll accounts are not public. 
What we do know is that this is not the complete set of active Russian trolls, because 6 days prior to this writing Twitter announced they have discovered over 1K more troll accounts.\footnote{\url{https://blog.twitter.com/official/en_us/topics/company/2018/2016-election-update.html}}
Nonetheless, it constitutes an invaluable ``ground truth'' dataset enabling efforts to shed light on the behavior of state-sponsored troll accounts.

\descr{Baseline dataset.} We also compile a list of random Twitter users, while ensuring that the distribution of the average number of tweets per day posted by the random users is similar to the one by trolls.
To calculate the average number of tweets posted by an account, we find the first tweet posted after January 1, 2016 and retrieve the overall tweet count. %
This number is then divided by the number of days since account creation.
Having selected a set of 1K random users, we then collect all their tweets between January 2016 and September 2017, obtaining a total of 96K tweets.
We follow this approach as it gives a good approximation of posting behavior, even though
it might not be perfect, since (1) Twitter accounts can become more or less active over time, and (2) our datasets are based on the 1\% Streaming API, thus, we are unable to control the number of tweets we obtain for each account.

\begin{figure}[t]
\center
\subfigure[Hour of Day]{\includegraphics[width=0.49\columnwidth,height=1.1in]{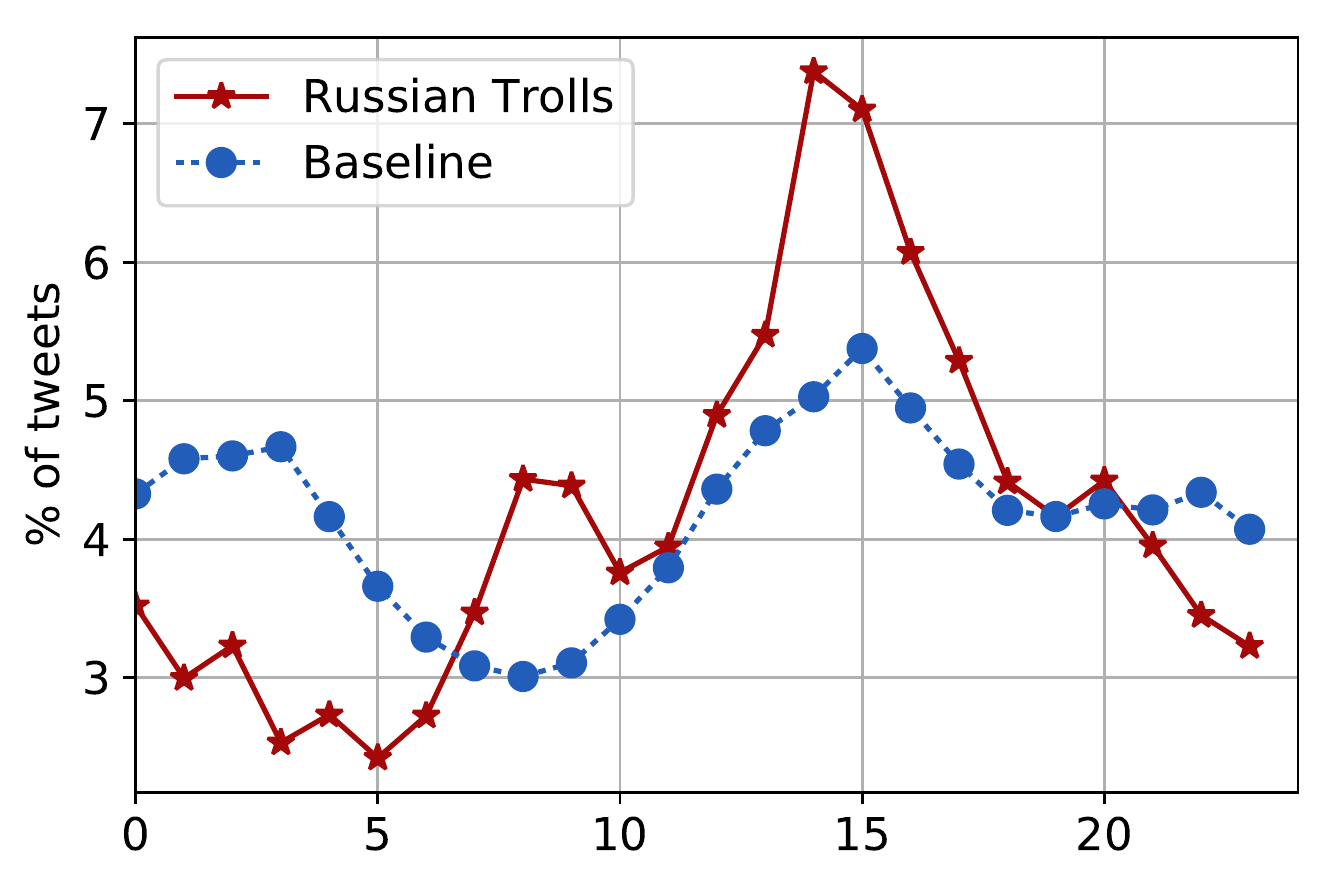}\label{subfig:counts_per_hour_day}}
\subfigure[Hour of Week]{\includegraphics[width=0.49\columnwidth,height=1.1in]{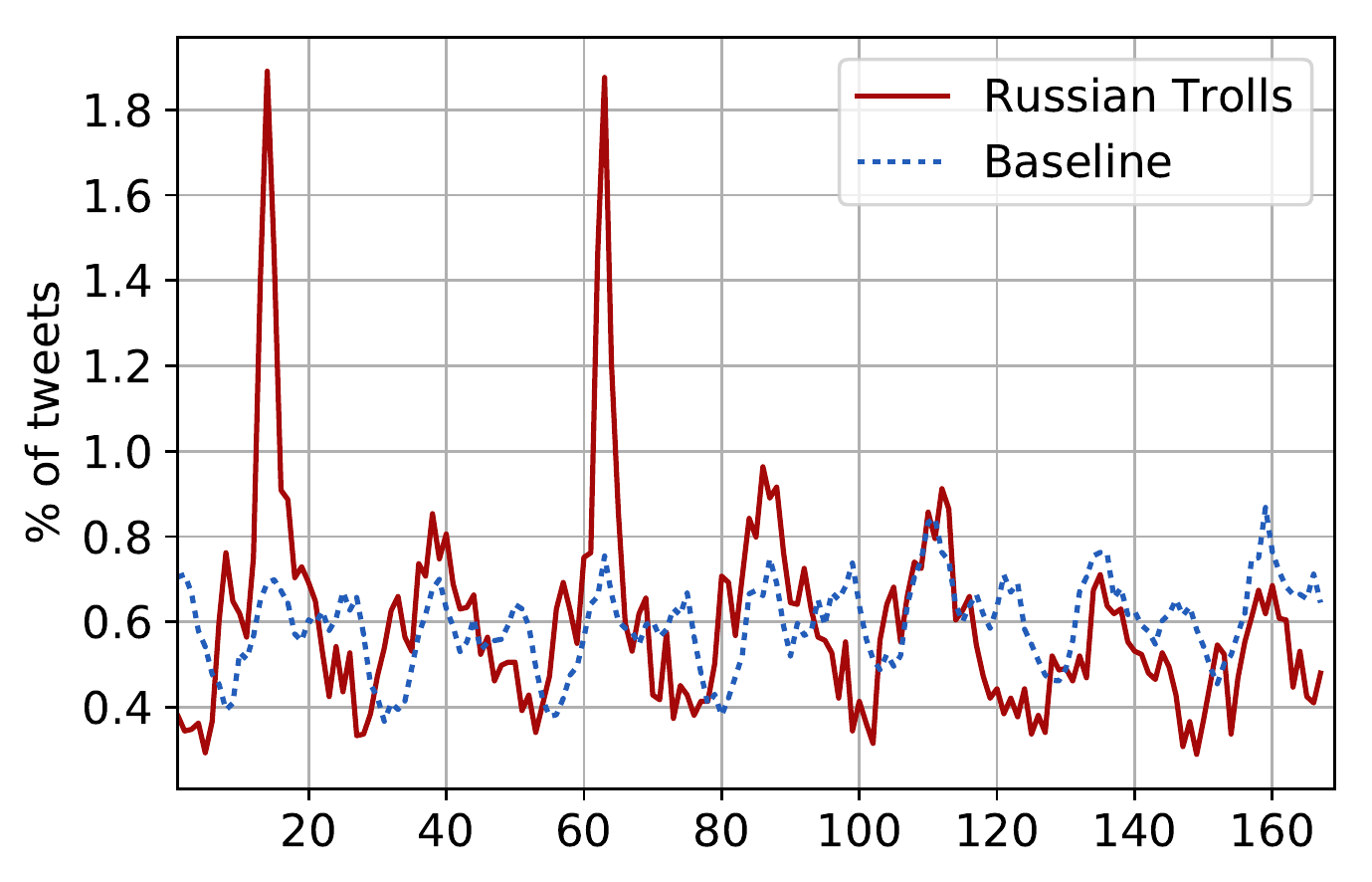}\label{subfig:counts_per_hour_week}}
  \caption{Temporal characteristics of tweets from Russian trolls and random Twitter users. }
\label{fig:temporal_analysis}
\end{figure}

\section{Analysis} \label{sec:analysis}
\longVer{
In this section, we present an in-depth analysis of the activities and the behavior of Russian trolls.
First, we provide a general characterization of the accounts and a geographical analysis of the locations they report.
Then, we analyze the content they share and how they evolved until their suspension by Twitter.
Finally, we present a case study of one specific account.
}
\longVer{
\subsection{General Characterization} \label{sec:general}
}
\descr{Temporal analysis.}
We observe that Russian trolls are continuously active on Twitter between January, 2016 and September, 2017, with a peak of activity just before the second US presidential debate (October 9, 2016).
Fig.~\ref{subfig:counts_per_hour_day} shows that most tweets from the trolls are posted between 14:00 and 15:00 UTC.
In Fig.~\ref{subfig:counts_per_hour_week}, we also report temporal characteristics based on hour of the week,
finding that both datasets follow a diurnal pattern, while trolls' activity peaks around 14:00 and 15:00 UTC on Mondays and Wednesdays.
Considering that Moscow is three hours ahead UTC, this distribution does not rule out that tweets might actually be posted from Russia.

\begin{figure}[t!]
\centering
\includegraphics[width=0.7\columnwidth]{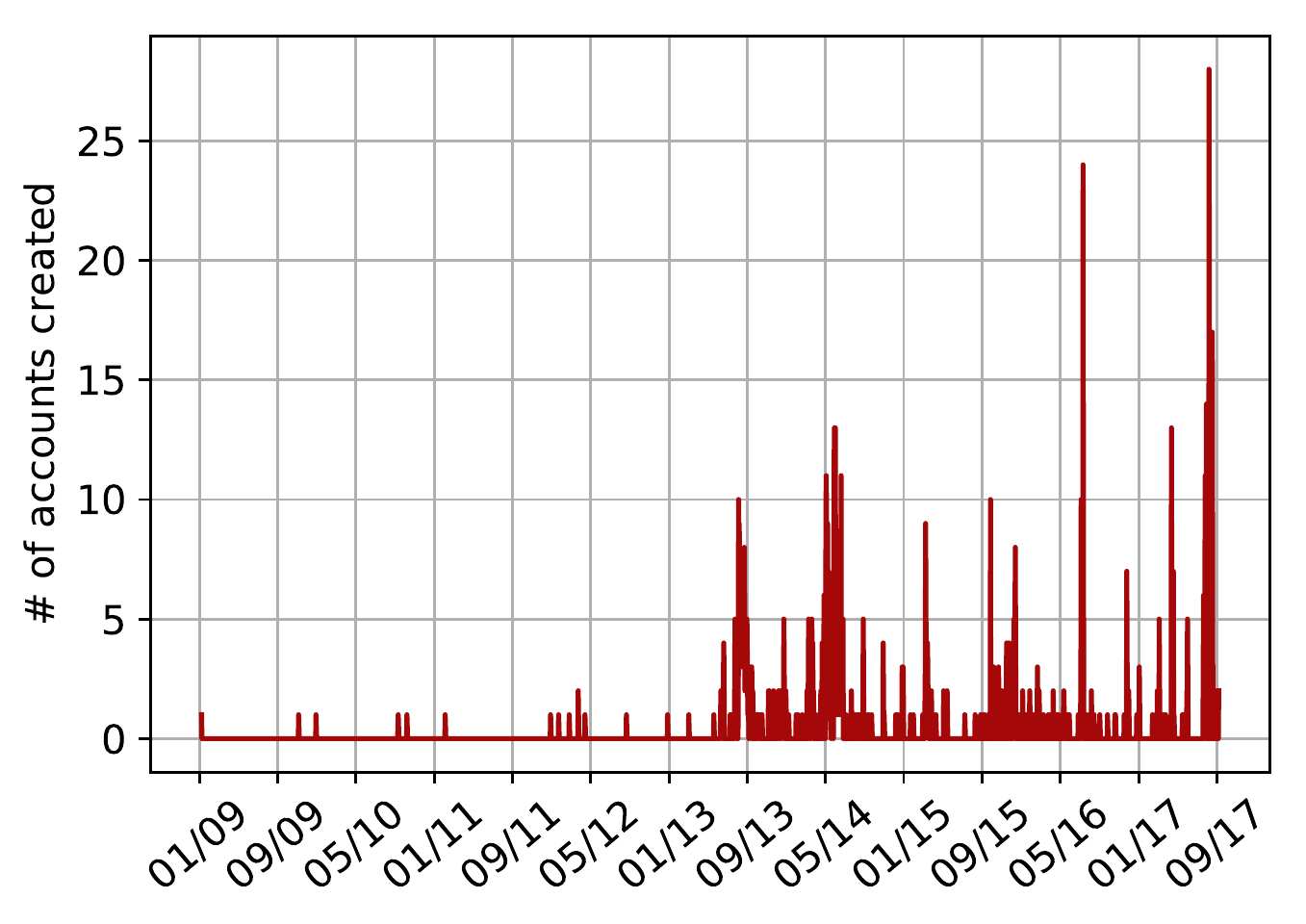}
\caption{Number of Russian troll accounts created per day.}
\label{fig:counts_created}
\end{figure}

\descr{Account creation.}  Next, we examine the dates when the trolls infiltrated Twitter,
by looking at the account creation dates.
From Fig.~\ref{fig:counts_created}, %
we observe that 71\% of them are actually created before 2016.
There are some interesting peaks, during 2016 and 2017: for instance, 24 accounts are created on July 12, 2016, approx. a week before the Republican National Convention (when Donald Trump received the nomination), %
while 28 appear on August 8, 2017, a few days before the infamous Unite the Right rally in Charlottesville. %
Taken together, this might be evidence of coordinated activities aimed at manipulating users' opinions with respect to specific events.

\descr{Account characteristics.}
We also shed light on the troll account profile information.
In Table~\ref{tbl:account_desc}, we report the top ten words appearing in the screen names and the descriptions of Russian trolls, as well as character 4-grams for screen names and word bigrams for profile descriptions.
Interestingly, a substantial number of Russian trolls pose as news outlets, evident from the use of the term ``news'' in both the screen name (1.3\%) and the description (10.7\%).
Also, it seems they attempt to increase the number of their followers, thus their reach of Twitter users, by nudging users to follow them (see, e.g., ``follow me'' appearing in almost 8\% of the accounts).
Finally, 10.3\% of the Russian trolls describe themselves as Trump supporters: ``trump'' and ``maga'' (Make America Great Again, one of Trump campaign's main slogans).

\begin{table}[]
\centering
\resizebox{0.99\columnwidth}{!}{
\setlength{\tabcolsep}{0.4em} %
\begin{tabular}{@{}lrlrlrlr@{}}
\toprule
\multicolumn{4}{c}{\textbf{Screen Name}}  & \multicolumn{4}{c}{\textbf{Description}} \\ %
\textbf{Word} & \textbf{(\%)} & \textbf{4-gram} & \textbf{(\%)} & \textbf{Word} & \textbf({\%)} & \textbf{Word bigram} & \textbf{(\%)} \\ \midrule
news                 & 1.3\%         & news                        & \multicolumn{1}{r|}{1.5\%}       & news                  & 10.7\%        & follow me                    & 7.8\%         \\
bote                 & 1.2\%         & line                        & \multicolumn{1}{r|}{1.5\%}       & follow                & 10.7\%        & breaking news                & 2.6\%         \\
online               & 1.1\%         & blac                        & \multicolumn{1}{r|}{1.3\%}       & conservative          & 8.1\%         & news aus                     & 2.1\%         \\
daily                & 0.8\%         & bote                        & \multicolumn{1}{r|}{1.3\%}       & trump                 & 7.8\%         & uns in                       & 2.1\%         \\
today                & 0.6\%         & rist                        & \multicolumn{1}{r|}{1.1\%}       & und                   & 6.2\%         & deiner stdt                  & 2.1\%         \\
ezekiel2517          & 0.6\%         & nlin                        & \multicolumn{1}{r|}{1.1\%}       & maga                  & 5.9\%         & die news                     & 2.1\%         \\
maria                & 0.5\%         & onli                        & \multicolumn{1}{r|}{1.0\%}       & love                  & 5.8\%         & wichtige und                 & 2.1\%         \\
black                & 0.5\%         & lack                        & \multicolumn{1}{r|}{1.0\%}       & us                    & 5.3\%         & nachrichten aus              & 2.1\%         \\
voice                & 0.4\%         & bert                        & \multicolumn{1}{r|}{1.0\%}       & die                   & 5.0\%         & aus deiner                   & 2.1\%         \\
martin               & 0.4\%         & poli                        & \multicolumn{1}{l|}{1.0\%}        & nachrichten           & 4.3\%         & die dn                       & 2.1\%         \\ \bottomrule
\end{tabular}
}
\caption{Top 10 words found in Russian troll screen names and account descriptions. We also report character 4-grams for the screen names and word bigrams for the description.}
\label{tbl:account_desc}
\end{table}

\begin{figure}[t!]
\center
\subfigure[]{\includegraphics[width=0.49\columnwidth]{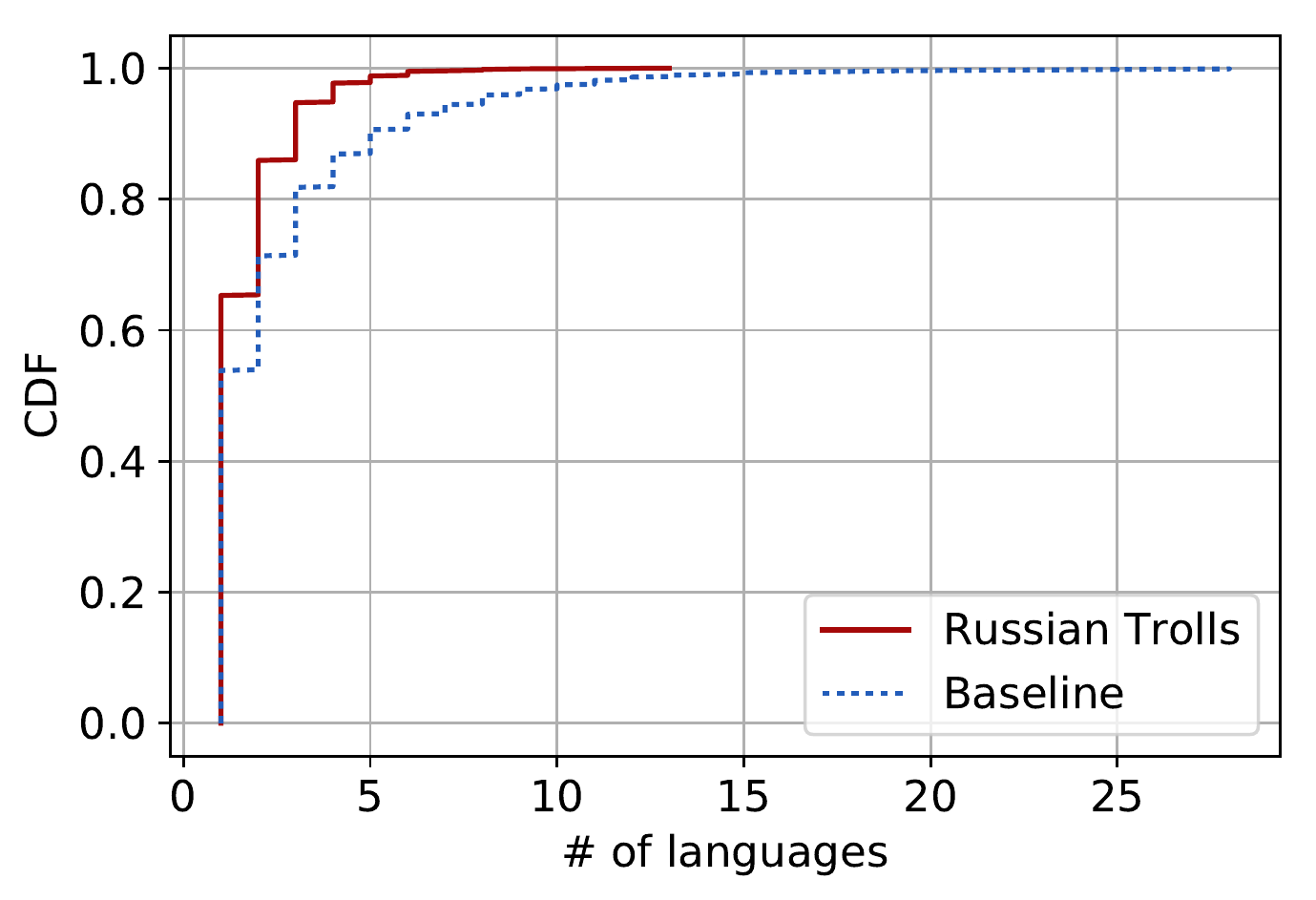}\label{subfig:cdf_languages_user}}
\subfigure[]{\includegraphics[width=0.49\columnwidth]{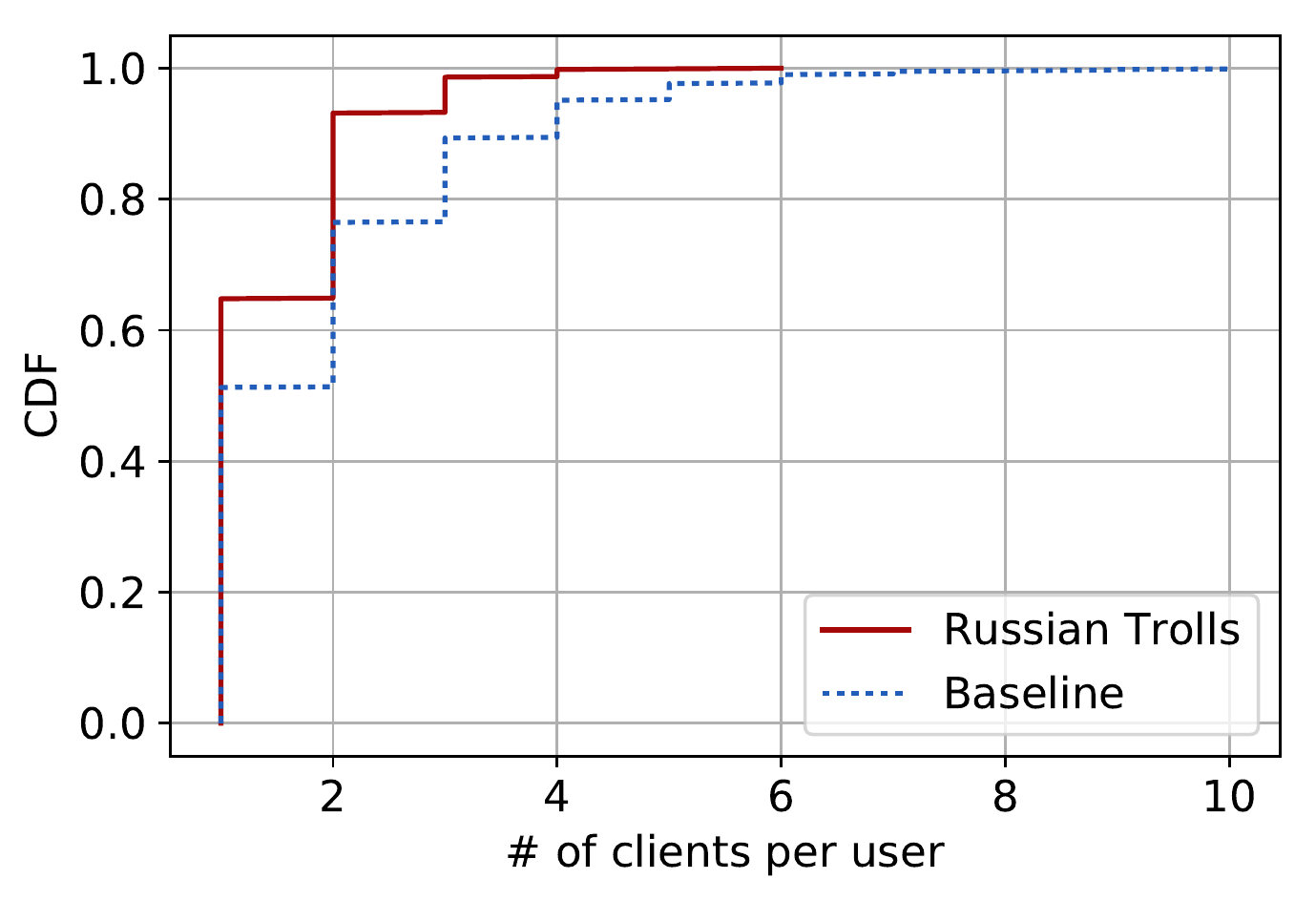}\label{subfig:cdf_sources_user}}
  \caption{CDF of number of (a) languages used (b) clients used per user. }
\label{fig:cdf_lang_sources}
\end{figure}

\descr{Language.} Looking at the language (as provided via the Twitter API) of the tweets posted by Russian trolls, we find that most of them (61\%) are in English, although a substantial portion are in Russian (27\%), and to a lesser extent in German (3.5\%).
In Fig.~\ref{subfig:cdf_languages_user}, we plot the cumulative distribution function (CDF) of the number of different languages for each user: 64\% of the Russian trolls post all their tweets in only one language,
compared to only 54\% for random users.
Overall, by comparing the two distributions, we observe that random Twitter users tend to use more languages in their tweets compared to the trolls.

\begin{table}[t]
\centering
\resizebox{0.9\columnwidth}{!}{
\begin{tabular}{lr|lr}
\toprule
\textbf{Client (Trolls)} & \multicolumn{1}{r}{\textbf{(\%)}} & \textbf{Client (Baseline)} & \textbf{(\%)} \\ \midrule
Twitter Web Client               & 50.1\%       & TweetDeck          & 32.6\%        \\
twitterfeed                      & 13.4\%        & Twitter for iPhone        & 26.2\%        \\
Twibble.io                            & 9.0\%         & Twitter for Android         & 22.6\%        \\
IFTTT                      & 8.6\%         &   Twitter Web Client               & 6.1\%         \\
TweetDeck                        & 8.3\%         & GrabInbox               & 2.0\%         \\
NovaPress                        & 4.6\%         & Twitter for iPad           & 1.4\%         \\
dlvr.it                          & 2.3\%         & IFTTT                      & 1.0\%         \\
Twitter for iPhone               & 0.8\%         & twittbot.net                   & 0.9\%         \\
Zapier.com                       & 0.6\%         & Twitter for BlackBerry              & 0.6\%         \\
Twitter for Android              & 0.6\%         & Mobile Web (M2)            & 0.4\%         \\ \bottomrule
\end{tabular}
}
\caption{Top 10 Twitter clients (as \% of tweets).}
\label{tbl:top_sources}
\end{table}

\descr{Client.} Finally, we analyze the clients used to post tweets.
We do so since previous work~\cite{egele2017towards} shows that the client used by official or professional accounts are quite different that the ones used by regular users.
Table~\ref{tbl:top_sources} reports the top 10 clients for both Russian trolls and baseline users.
We find the latter prefer to use Twitter clients for mobile devices (48\%) and the TweetDeck dashboard (32\%), whereas, the former mainly use the Web client (50\%).
We also assess how many different clients Russian trolls use throughout our dataset:
in Fig.~\ref{subfig:cdf_sources_user}, we plot the CDF of the number of clients used per user.
We find that 65\%  of the Russian trolls use only one client, 28\% of them two different clients, and the rest more than three, which is overall less than the random baseline users.

\shortVer{
\begin{figure}[t]
\centering
\includegraphics[width=0.85\columnwidth]{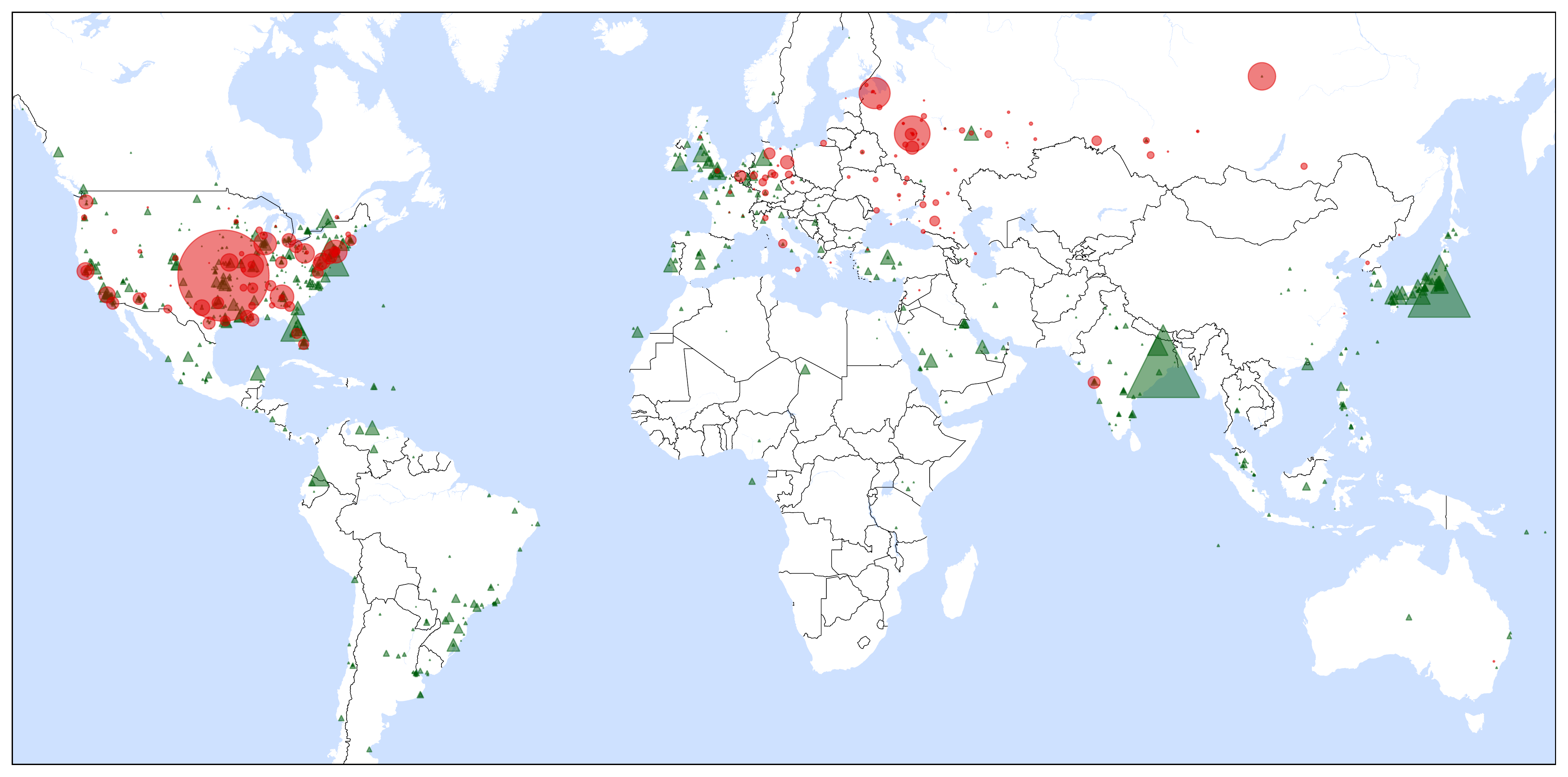}
   \caption{Distribution of reported locations for tweets by Russian trolls (red circles) and baseline (green triangles).}
\label{fig:locations_map}
\end{figure}
}

\longVer{
\begin{figure*}[t]
\centering
\includegraphics[width=0.7\textwidth]{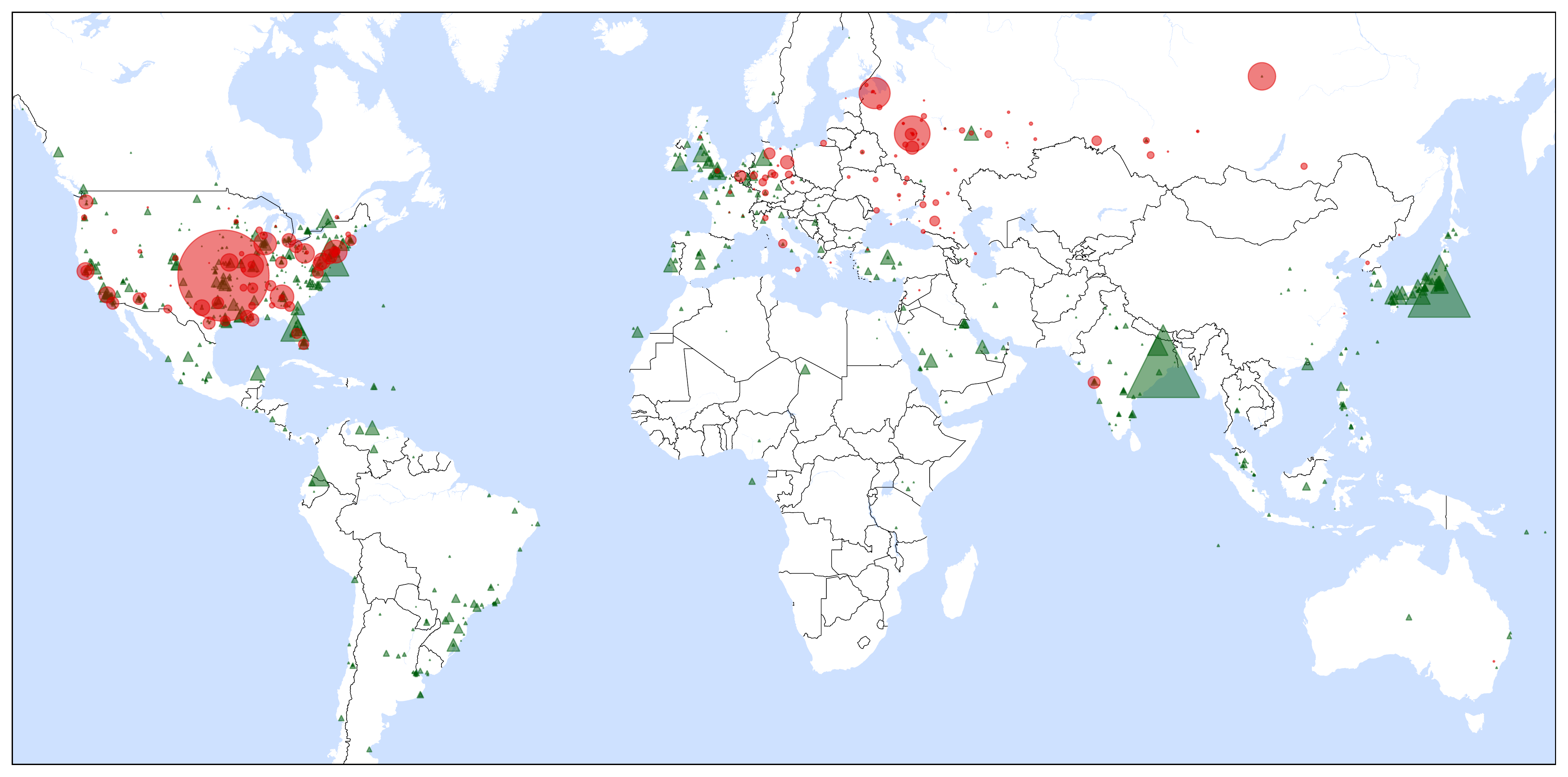}
   \caption{Distribution of reported locations for tweets by Russian trolls (red circles) and baseline (green triangles).}
\label{fig:locations_map}
\end{figure*}
}

\longVer{
\subsection{Geographical Analysis}
}
\descr{Location.} We then study users' location, relying on the self-reported location field in their profiles.
Note that users not only may leave it empty, but also change it any time they like,
so we look at locations for each tweet.
We retrieve it for 75\% of the tweets by Russian trolls, gathering 261 different entries, which we convert to a physical location using the Google Maps Geocoding API. %
In the end, we obtain 178 unique locations for the trolls,
as depicted in Fig.~\ref{fig:locations_map} (red circles).
The size of the circles on the map indicates the number of tweets that appear at each location.
We do the same for the baseline, getting 2,037 different entries, converted by the API to 894 unique locations. %
We observe that most of the tweets from Russian trolls come from locations within the USA and Russia, and some from European countries, like Germany, Belgium, and Italy.
On the other hand, tweets in our baseline are more uniformly distributed across the globe, with many tweets from North and South America, Europe, and Asia.
This suggests that Russian trolls may be pretending to be from certain countries, e.g., USA or Germany, aiming to pose as locals and better manipulate opinions.
This explanation becomes more plausible when we consider that a plurality of trolls' tweets have their location set as a generic form of ``US,'' as opposed to a specific city, state, or even region.
Interestingly, the 2nd, 3rd, and 4th most popular location for trolls to tweet from are Moscow, St. Petersburg, and a generic form of ``Russia.''
We also assess whether users change their country of origin based on the self-reported location:
only a negligible percentage (1\%) of trolls change their country, while for the baseline the percentage is 16\%.

\longVer{
\begin{figure}[t]
\center
\subfigure[]{\includegraphics[width=0.49\columnwidth]{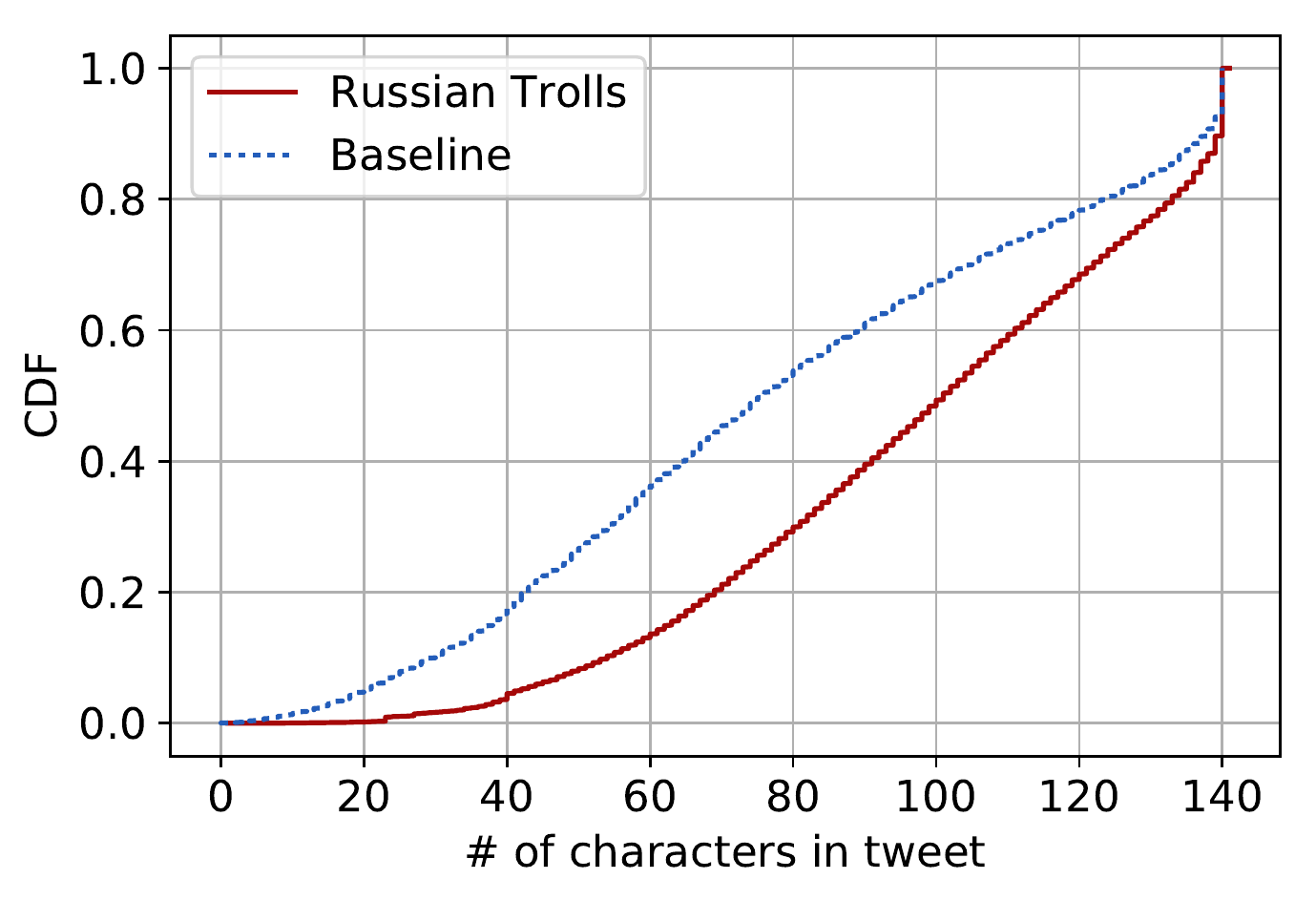}\label{subfig:cdf_characters}}
\subfigure[]{\includegraphics[width=0.49\columnwidth]{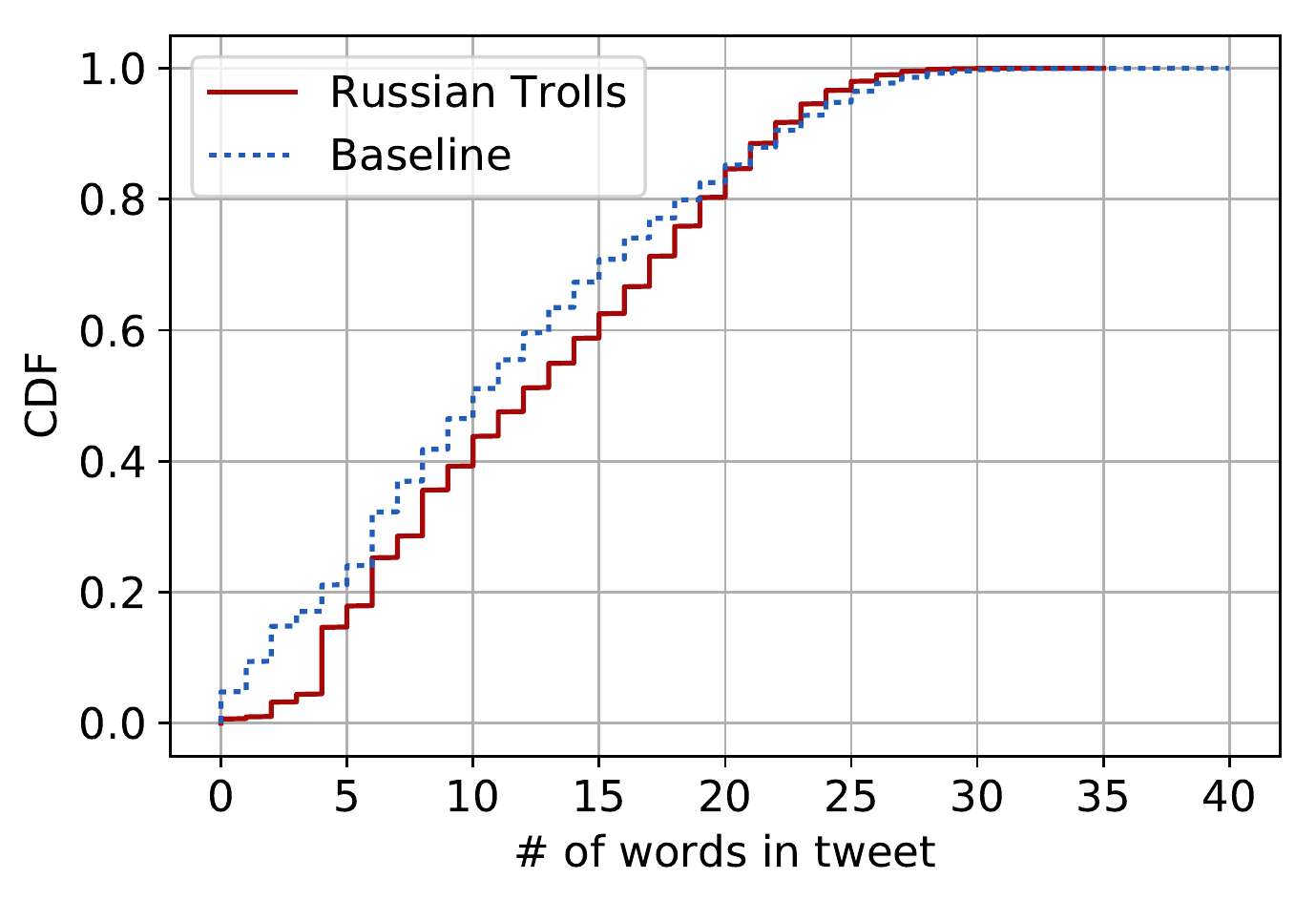}\label{subfig:cdf_words}}
  \caption{CDF of the number of (a) characters and (b) words in each tweet.}
\label{fig:cdfs_text}
\end{figure}

\begin{table}[t]
\centering
\resizebox{0.9\columnwidth}{!}{
\begin{tabular}{lrlr}
\toprule
\textbf{Timezone (Trolls)}           & \multicolumn{1}{r}{\textbf{(\%)}}                    & \textbf{Timezone (Baseline)}            & \textbf{(\%)} \\ \midrule
Eastern Time  & \multicolumn{1}{r|}{38.87\%} & Athens                       & 24.41\%        \\
Pacific Time  & \multicolumn{1}{r|}{18.37\%} & Pacific Time                    & 21.41\%        \\
Volgograd                   & \multicolumn{1}{r|}{10.03\%} & London                       & 21.27\%        \\
Central Time  & \multicolumn{1}{r|}{9.43\%}  & Tokyo               & 3.83\%        \\
Moscow                      & \multicolumn{1}{r|}{8.18\%}  & Central Time                      & 3.75\%        \\
Bern                        & \multicolumn{1}{r|}{2.56\%}  & Eastern Time                      & 2.10\%        \\
Minsk                       & \multicolumn{1}{r|}{2.06\%}  & Seoul                  & 1.97\%        \\
Yerevan                     & \multicolumn{1}{r|}{1.96\%}  & Brasilia                       & 1.97\%        \\
Nairobi                     & \multicolumn{1}{r|}{1.52\%}  & Buenos Aires                         & 1.92\%        \\
Baku                        & \multicolumn{1}{r|}{1.29\%}  & Urumqi  & 1.50\%        \\ \bottomrule
\end{tabular}
}
\caption{Top 10 timezones (as \% of tweets).}
\label{tbl:top_timezones}
\end{table}

\descr{Timezone.} We then study the timezone chosen by the users in their account setting.
In Table~\ref{tbl:top_timezones}, we report the top 10 timezones for each dataset, in terms of the corresponding tweet volumes.
Two thirds of the tweets by trolls appear to be from US timezones, while a substantial percentage (18\%) from Russian ones. %
Whereas, the baseline has a more diverse set of timezones, which seems to mirror findings from our location analysis.

We also check whether users change their timezone settings,
finding that 7\% of the Russian trolls do so two to three times.
The most popular changes are Berlin to Bern (18 times), Nairobi to Moscow (10), and Nairobi to Volgograd (10).%
By contrast, this almost never happens for baseline accounts.
}

\longVer{
\subsection{Content Analysis}
\descr{Text.} Next, we quantify the number of characters and words contained in each tweet,
and plot the corresponding CDF in Fig.~\ref{fig:cdfs_text}, finding that Russian trolls tend to post longer tweets.
}

\descr{Media.} We then assess whether Russian trolls use images and videos in a different way than random baseline users.
For Russian trolls (resp., baseline accounts), 66\% (resp., 73\%) of the tweets include no images, 32\% (resp., 18\%) exactly one image, and 2\% (resp., 9\%) more than one.
This suggests that Russian trolls disseminate a considerable amount of information via single-image tweets.
As for videos, only 1.5\% of the tweets from Russian trolls includes a video, as opposed to 6.4\% for baseline accounts.

\begin{table}[]
\centering
\setlength{\tabcolsep}{0.2em} %
\hspace*{-0.2cm}
\resizebox{\columnwidth}{!}{
\begin{tabular}{lrlrlrlr}
\toprule
\multicolumn{4}{c}{\textbf{Trolls}}                                       & \multicolumn{4}{c}{\textbf{Baseline}}       \\
\textbf{Hashtag}          & \textbf{(\%)}  & \textbf{Hashtag}           & \multicolumn{1}{l}{\textbf{(\%)}}   & \textbf{Hashtag}      & \textbf{(\%)}  & \textbf{Hashtag}      & \textbf{(\%)}  \\ \midrule
news             & 7.2\% & US                & \multicolumn{1}{r|}{0.7\%} & iHeartAwards & 1.8\% & UrbanAttires & 0.6\% \\
politics         & 2.6\% & tcot              & \multicolumn{1}{r|}{0.6\%} & BestFanArmy  & 1.6\% & Vacature     & 0.6\% \\
sports           & 2.1\% & PJNET             & \multicolumn{1}{r|}{0.6\%} & Harmonizers  & 1.0\% & mPlusPlaces  & 0.6\% \\
business         & 1.4\% & entertainment     & \multicolumn{1}{r|}{0.5\%} & iOSApp       & 0.9\% & job          & 0.5\% \\
money            & 1.3\% & top               & \multicolumn{1}{r|}{0.5\%} & JouwBaan     & 0.9\% & Directioners & 0.5\% \\
world            & 1.2\% & topNews           & \multicolumn{1}{r|}{0.5\%} & vacature     & 0.9\% & JIMIN        & 0.5\% \\
MAGA             & 0.8\% & ISIS              & \multicolumn{1}{r|}{0.4\%} & KCA          & 0.9\% & PRODUCE101   & 0.5\% \\
health           & 0.8\% & Merkelmussbleiben & \multicolumn{1}{r|}{0.4\%} & Psychic      & 0.8\% & VoteMainFPP  & 0.5\% \\
local            & 0.7\% & IslamKills        & \multicolumn{1}{r|}{0.4\%} & RT           & 0.8\% & Werk         & 0.4\% \\
BlackLivesMatter & 0.7\% & breaking          & \multicolumn{1}{l|}{0.4\%} & Libertad2016 & 0.6\% & dts          & 0.4\% \\ \bottomrule
\end{tabular}
}
\caption{Top 20 hashtags in tweets from Russian trolls and baseline users.}
\label{tbl:top_hashtags}
\end{table}

\descr{Hashtags.} Our next step is to study the use of hashtags in tweets.
Russian trolls use at least one hashtag in 32\% of their tweets, compared to 10\% for the baseline.
Overall, we find 4.3K and 7.1K unique hashtags for trolls and random users, respectively,
with 74\% and 78\% of them only appearing once.
In Table~\ref{tbl:top_hashtags}, we report the top 20 hashtags for both datasets.
Trolls appear to use hashtags to disseminate news (7.2\%) and politics (2.6\%) related content, but also use several that might be indicators of propaganda and/or controversial topics, e.g., \#ISIS, \#IslamKills, and \#BlackLivesMatter.
For instance, we find some notable examples including: 
``We just have to close the borders, `refugees' are simple terrorists \#IslamKills'' on March 22, 2016,
``\#SyrianRefugees ARE TERRORISTS from \#ISIS \#IslamKills'' on March 22, 2016, and
``WATCH: Here is a typical \#BlackLivesMatter protester:  `I hope I kill all white babes!' \#BatonRouge $<$url$>$'' on July 17, 2016. 

We also study when these hashtags are used by the trolls, finding that most of them are well distributed over time. However, there are some interesting exceptions, e.g., with \#Merkelmussbleiben (a hashtag seemingly supporting Angela Merkel) and \#IslamKills. Specifically, tweets with the former appear exclusively on July 21, 2016,
while the latter on March 22, 2016, when a terrorist attack took place at Brussels airport.
These two examples illustrate how the trolls may be coordinating to push specific narratives on Twitter.

\descr{Mentions.} We find that 46\% of trolls' tweets include {\em mentions} %
 to 8.5K unique Twitter users.
This percentage is much higher for the random baseline users (80\%, to 41K users).
Table~\ref{tbl:top_mentions} reports the 20 top mentions we find in tweets from Russian trolls and baseline users.
We find several Russian accounts, like `leprasorium' (a popular Russian account that mainly posts memes) in 2\% of the mentions, as well as popular politicians like `realDonaldTrump' (0.6\%).
The practice of mentioning politicians on Twitter may reflect an underlying strategy to mutate users' opinions regarding a particular political topic, which has been also studied in previous work~\cite{conover2011political}.

\begin{table}[]
\centering
\setlength{\tabcolsep}{0.2em} %
\hspace*{-0.2cm}
\resizebox{1.06\columnwidth}{!}{
\begin{tabular}{lrlrlrlr}
\hline
\multicolumn{4}{c}{\textbf{Trolls}}                                                     & \multicolumn{4}{c}{\textbf{Baseline}}                               \\
\textbf{Mention} & \textbf{(\%)} & \textbf{Mention} & \multicolumn{1}{l}{\textbf{(\%)}} & \textbf{Mention} & \textbf{(\%)} & \textbf{Mention} & \textbf{(\%)} \\ \hline
leprasorium      & 2.1\%         & postsovet        & \multicolumn{1}{r|}{0.4\%}        & TasbihIstighfar  & 0.3\%         & RasSpotlights    & 0.1\%         \\
zubovnik         & 0.8\%         & DLGreez          & \multicolumn{1}{r|}{0.4\%}        & raspotlights     & 0.2\%         & GenderReveals    & 0.1\%         \\
realDonaldTrump  & 0.6\%         & DanaGeezus       & \multicolumn{1}{r|}{0.4\%}        & FunnyBrawls      & 0.2\%         & TattedChanel     & 0.1\%         \\
midnight         & 0.6\%         & ruopentwit       & \multicolumn{1}{r|}{0.3\%}        & YouTube          & 0.2\%         & gemvius          & 0.1\%         \\
blicqer          & 0.6\%         & Spoontamer       & \multicolumn{1}{r|}{0.3\%}        & Harry\_Styles    & 0.2\%         & DrizzyNYC\_\_    & 0.1\%         \\
gloed\_up        & 0.6\%         & YouTube          & \multicolumn{1}{r|}{0.3\%}        & shortdancevids   & 0.2\%         & August\_Alsina\_ & 0.1\%         \\
wylsacom         & 0.5\%         & ChrixMorgan      & \multicolumn{1}{r|}{0.3\%}        & UrbanAttires     & 0.2\%         & RihannaBITCH\_   & 0.1\%         \\
TalibKweli       & 0.4\%         & sergeylazarev    & \multicolumn{1}{r|}{0.3\%}        & BTS\_twt         & 0.2\%         & sexualfeed       & 0.1\%         \\
zvezdanews       & 0.4\%         & RT\_com          & \multicolumn{1}{r|}{0.3\%}        & KylieJenner\_NYC & 0.2\%         & PetsEvery30      & 0.1\%         \\
GiselleEvns      & 0.4\%         & kozheed          & \multicolumn{1}{l|}{0.3\%}        & BaddiessNation   & 0.2\%         & IGGYAZALEAoO    & 0.1\%         \\ \hline
\end{tabular}
}
\caption{Top 20 mentions in tweets from trolls and baseline.}
\label{tbl:top_mentions}
\end{table}

\begin{figure}[t]
\centering
\includegraphics[width=0.7\columnwidth]{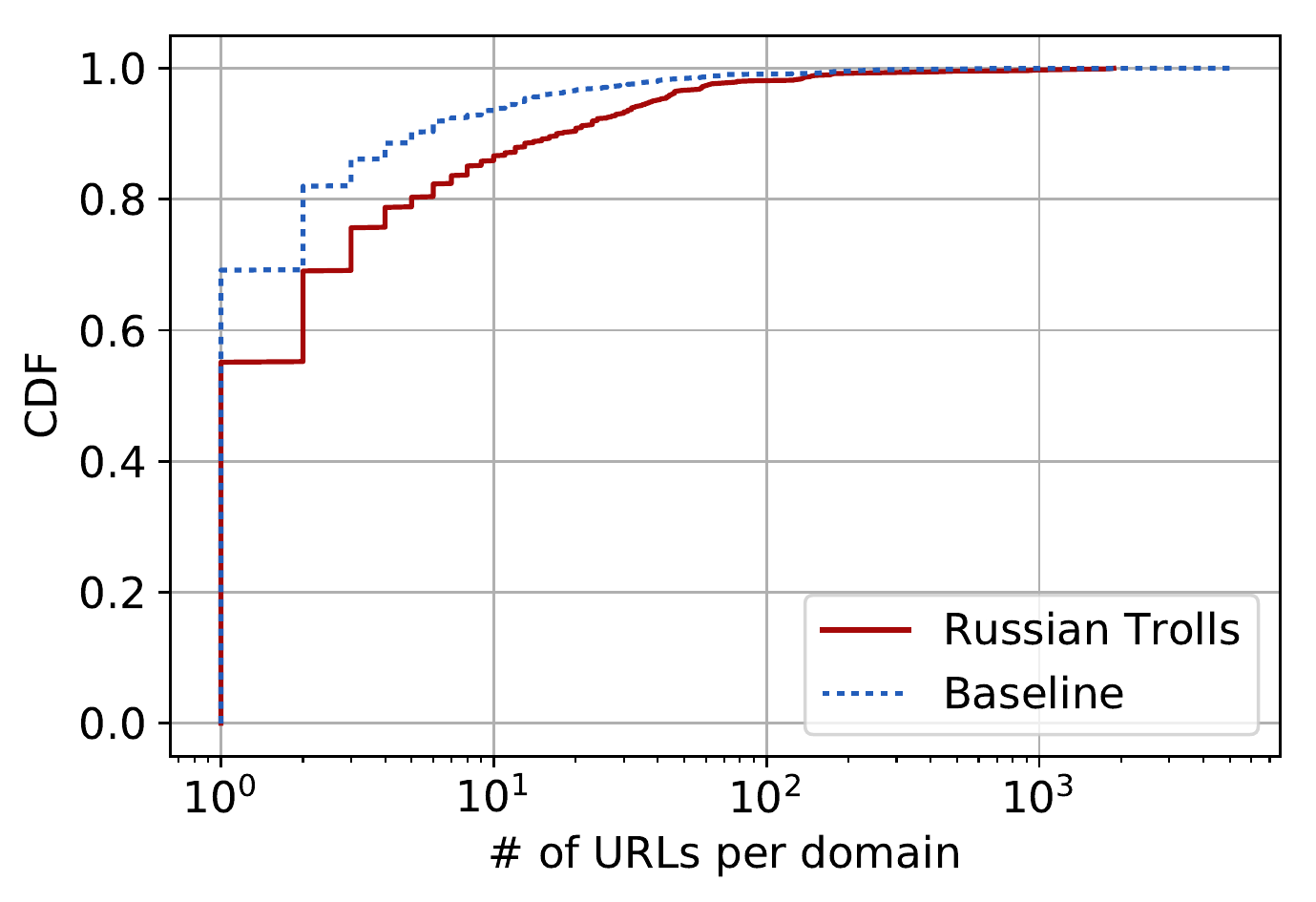}
  \caption{CDF of number of URLs per domain.}
\label{fig:cdf_domain_count}
\end{figure}

\descr{URLs.} We then analyze the URLs included in the tweets.
First of all, we note that 53\% of the trolls' tweets include at least a URL, compared to only 27\% for the random baseline.
There is an extensive presence of URL shorteners for both datasets, e.g., \url{bit.ly} (12\% for trolls and 26\% for the baseline) and \url{ift.tt} (10\% for trolls and 2\% for the baseline), therefore, in November 2017, we visit each URL to obtain the final URL after all redirections.
In Fig.~\ref{fig:cdf_domain_count}, we plot the CDF of the number of URLs per unique domain.
We observe that Russian trolls disseminate more URLs in their tweets compared to the baseline.
This might indicate that Russian trolls include URLs to increase their credibility and positive user perception; indeed, \cite{gupta2012credibility} show that adding a URL in a tweet correlates with higher credibility scores.
Also, in Table~\ref{tbl:top_domains}, we report the top 20 domains for both Russian trolls and the baseline.
Most URLs point to content within Twitter itself; 13\% and 35\%, respectively.
Links to a number of popular social networks like YouTube (1.8\% and 4.2\%, respectively) and Instagram (1.5\% and 1.9\%) appear in both datasets.
We also note that among the top 20 domains, there are also a number of news outlets linked from trolls' tweets, e.g., Washington Post (0.7\%), Seattle Times (0.7\%), and state-sponsored news outlets like RT (0.8\%) in trolls' tweets, but much less so from the baseline. %
\shortVer{
\begin{table}[t]
\centering
\footnotesize
\resizebox{0.75\columnwidth}{!}{%
\begin{tabular}{lr|lr}
\toprule
\textbf{Domain (Trolls)} & \multicolumn{1}{r}{(\%)} &\textbf{Domain (Baseline)}  & \textbf{ (\%)} \\ \midrule
twitter.com     & {12.81\%} & twitter.com    &  35.51\%\\
reportsecret.com                       & {7.02\%}  & youtube.com      & 4.21\%\\
riafan.ru          & {3.42\%}  & vine.co       & 3.94\%\\
politexpert.net      & {2.10\%}  & factissues.com      & 3.24\% \\
youtube.com      & {1.88\%}  & blogspot.com.cy     & 1.92\%      \\
vk.com               & {1.58\%}  & instagram.com       &   1.90\%  \\
instagram.com             & {1.53\%}  & facebook.com           & 1.68\% \\
yandex.ru                   & {1.50\%}  & worldstarr.info     &     1.47\%  \\
infreactor.org                     & {1.36\%}  & trendytopic.info         &  1.39\%\\
cbslocal.com    & {1.35\%}  & minibird.jp         & 1.25\%\\
 \bottomrule
\end{tabular}
}
\caption{Top 10 domains in tweets from trolls and the baseline.}
\label{tbl:top_domains}
\end{table}
}
\longVer{
\begin{table}[t]
\centering
\footnotesize
\resizebox{0.9\columnwidth}{!}{%
\begin{tabular}{lr|lr}
\toprule
\textbf{Domain (Trolls)} & \multicolumn{1}{r}{(\%)} &\textbf{Domain (Baseline)}  & \textbf{ (\%)} \\ \midrule
twitter.com     & {12.81\%} & twitter.com    &  35.51\%\\
reportsecret.com                       & {7.02\%}  & youtube.com      & 4.21\%\\
riafan.ru          & {3.42\%}  & vine.co       & 3.94\%\\
politexpert.net      & {2.10\%}  & factissues.com      & 3.24\% \\
youtube.com      & {1.88\%}  & blogspot.com.cy     & 1.92\%      \\
vk.com               & {1.58\%}  & instagram.com       &   1.90\%  \\
instagram.com             & {1.53\%}  & facebook.com           & 1.68\% \\
yandex.ru                   & {1.50\%}  & worldstarr.info     &     1.47\%  \\
infreactor.org                     & {1.36\%}  & trendytopic.info         &  1.39\%\\
cbslocal.com    & {1.35\%}  & minibird.jp         & 1.25\%\\
livejournal     & {1.35\%} & yaadlinksradio.com   &  1.24\%\\
nevnov.ru                       & {1.07\%}  & soundcloud.com      & 1.24\%\\
ksnt.com          & {1.01\%}  & linklist.me       & 1.15\%\\
kron4.com      & {0.93\%}  & twimg.com     & 1.09\% \\
viid.me      & {0.93\%}  & appparse.com     & 1.08\%      \\
newinform.com              & {0.89\%}  & cargobayy.net       &   0.88\%  \\
inforeactor.ru              & {0.84\%}  & virralclub.com          & 0.84\% \\
rt.com                & {0.81\%}  & tistory.com     &     0.50\%  \\
washigntonpost.com                      & {0.75\%}  & twitcasting.tv         &  0.49\%\\
seattletimes.com    & {0.73\%}  & nytimes.com          & 0.48\%\\ \bottomrule
\end{tabular}
}
\caption{Top 20 domains included in tweets from Russian trolls and baselines users.}
\label{tbl:top_domains}
\end{table}
}

\begin{figure}[t!]
\center
\subfigure[]{\includegraphics[width=0.49\columnwidth]{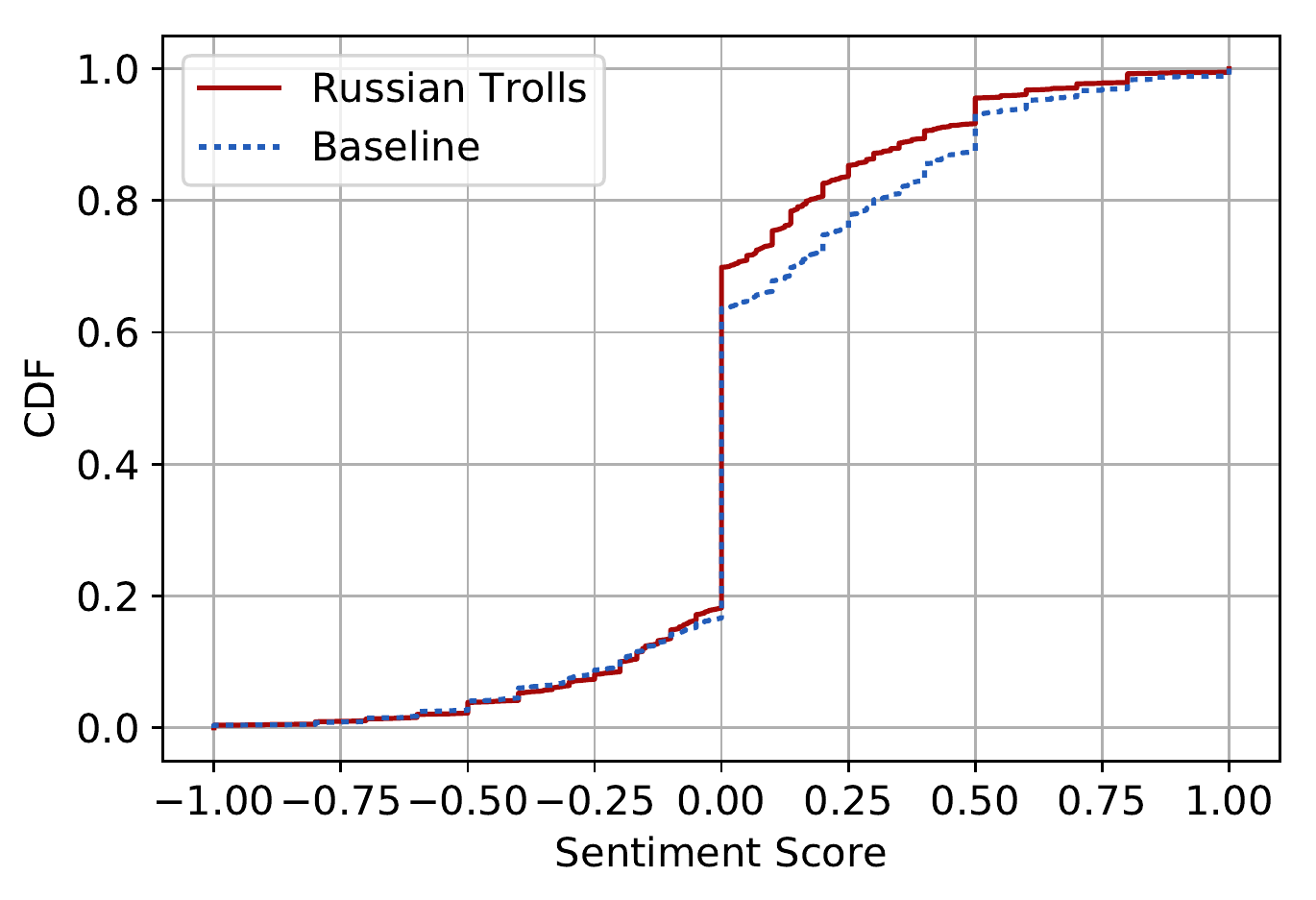}\label{subfig:cdf_sentiment}}
\subfigure[]{\includegraphics[width=0.49\columnwidth]{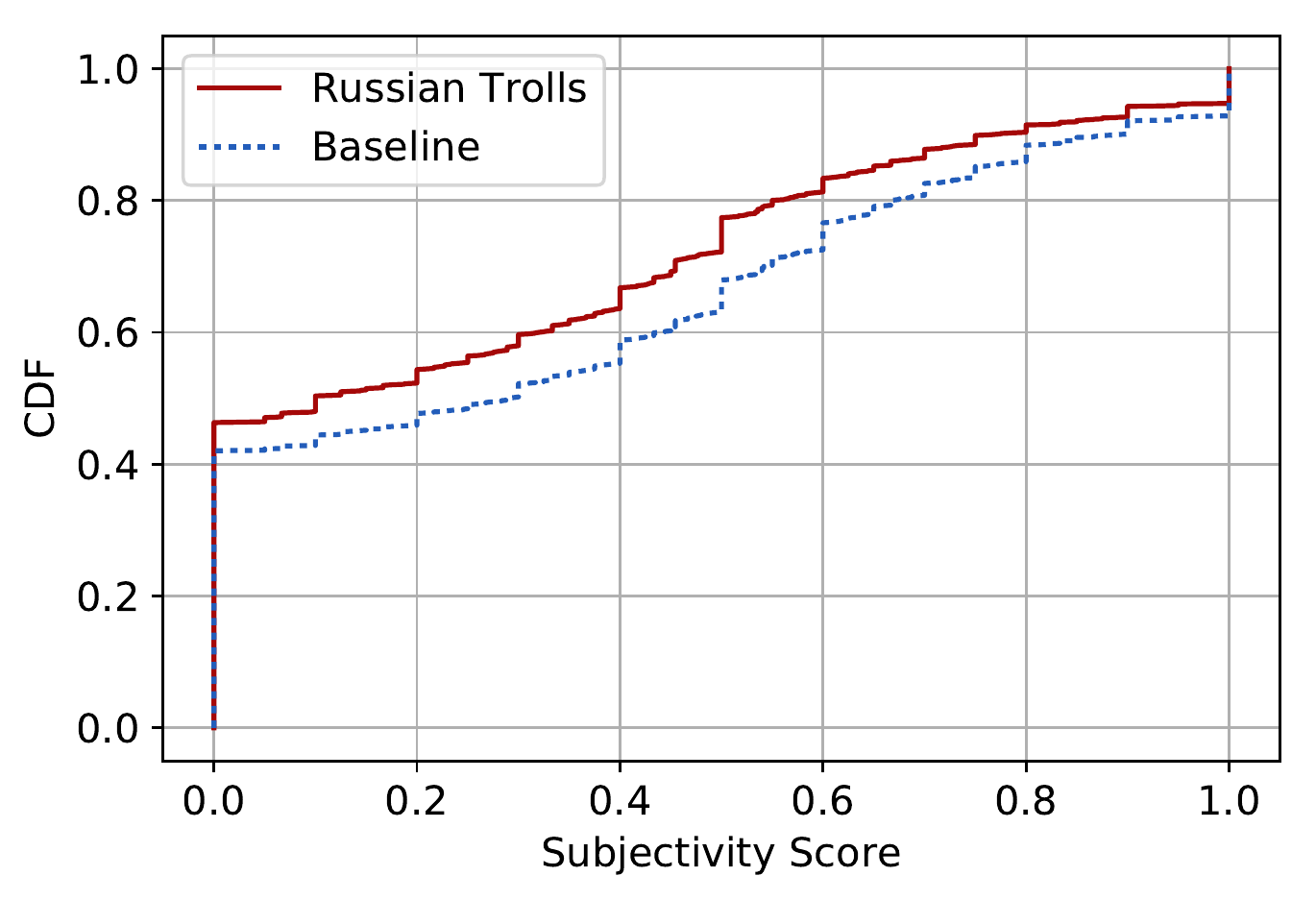}\label{subfig:cdf_subjectivity}}
\caption{CDF of sentiment and subjectivity scores for tweets of Russian trolls and random users.}
\label{fig:cdf_sentiment_subjectivity}
\end{figure}

\descr{Sentiment analysis.} Next, we assess the sentiment and subjectivity of each tweet for both datasets\shortVer{.}\longVer{
using the Pattern library~\cite{smedt2012pattern}.}
Fig.~\ref{subfig:cdf_sentiment} plots the CDF of the sentiment scores of tweets posted by Russian trolls and our baseline users.
We observe that 30\% of the tweets from Russian trolls have a positive sentiment, and 18\% negative.
These scores are not too distant from those of random users where 36\% are positive and 16\% negative, however, Russian trolls exhibit a unique behavior in terms of sentiment, as a two-sample Kolmogorov-Smirnov test unveils significant differences between the distributions ($p < 0.01$).
Overall, we observe that Russian trolls tend to be more negative/neutral, while our baseline is more positive.
We also compare  subjectivity scores (Fig.~\ref{subfig:cdf_subjectivity}), finding that  tweets from trolls tend to be more subjective; again, we perform significance tests revealing differences between the two distributions ($p < 0.01$).

\begin{table*}[]
\centering
\resizebox{0.9\textwidth}{!}{
\begin{tabular}{@{}rl|rl@{}}
\toprule
{\bf Topic} & \multicolumn{1}{l}{\textbf{Terms (Trolls)}} & {\textbf{Topic}} & \textbf{Terms (Baseline)}                                                               \\ \midrule
1  & trump, black, people, really, one, enlist, truth, work, can, get                 & 1           & want, can, just, follow, now, get, see, don, love, will                                 \\
2  & trump, year, old, just, run, obama, breaking, will, news, police                 & 2           & 2016, july, come, https, trump, social, just, media, jabberduck, get                    \\
3  & new, trump, just, breaking, obamacare, one, sessions, senate, politics, york     & 3           & happy, best, make, birthday, video, days, come, back, still, little                     \\
4  & man, police, news, killed, shot, shooting, woman, dead, breaking, death          & 4           & know, never, get, love, just, night, one, give, time, can                               \\
5  & trump, media, tcot, just, pjnet, war, like, video, post, hillary                 & 5           & just, can, everyone, think, get, white, fifth, veranomtv2016, harmony, friends          \\
6  & sports, video, game, music, isis, charlottesville, will, new, health, amb        & 6           & good, like, people, lol, don, just, look, today, said, keep                             \\
7  & can, don, people, want, know, see, black, get, just, like                        & 7           & summer, seconds, team, people, miss, don, will, photo, veranomtv2016, new               \\
8  & trump, clinton, politics, hillary, video, white, donald, president, house, calls & 8           & like, twitter, https, first, can, get, music, better, wait, really                      \\
9  & news, world, money, business, new, one, says, state, 2016, peace                 & 9           & dallas, right, fuck, vote, police, via, just, killed, teenchoice, aldubmainecelebration \\
10 & now, trump, north, korea, people, right, will, check, just, playing              & 10          & day, black, love, thank, great, new, now, matter, can, much                             \\ \bottomrule
\end{tabular}
}
\caption{Terms extracted from LDA topics of tweets from Russian trolls and baseline users.}
\label{tbl:lda_topics}
\end{table*}

\descr{LDA analysis.} %
We also use the Latent Dirichlet Allocation (LDA) model to analyze
tweets' semantics.
We train an LDA model for each of the datasets and extract 10 distinct topics with 10 words,
as reported in Table~\ref{tbl:lda_topics}.
Overall, topics from Russian trolls refer to specific world events (e.g., Charlottesville) as well as specific news related to politics (e.g., North Korea and Donald Trump).
By contrast, topics extracted from the random sample are more general %
(e.g., tweets regarding birthdays).

\longVer{
\subsection{Account Evolution}
}

\descr{Screen name changes. }
Previous work~\cite{mariconti2017s} has shown that malicious accounts often change their screen name in order to assume different identifies.
Therefore, we investigate whether trolls show a similar behavior, as they might  change the narrative with which they are attempting to influence public opinion.
Indeed, we find that 9\% of the accounts operated by trolls change their screen name, up to 4 times during the course of our dataset.
Some examples include changing screen names from ``OnlineHouston'' to ``HoustonTopNews,'' or  ``Jesus Quintin Perez'' to ``WorldNewsPolitics,''
in a clear attempt to pose as news-related accounts.
In our baseline, we find that 19\% of the accounts changed their Twitter screen names, up to 11 times during our dataset; highlighting that changing screen names is a common behavior of Twitter users in general.

\begin{figure}[t]
\center
\subfigure[]{\includegraphics[width=0.49\columnwidth]{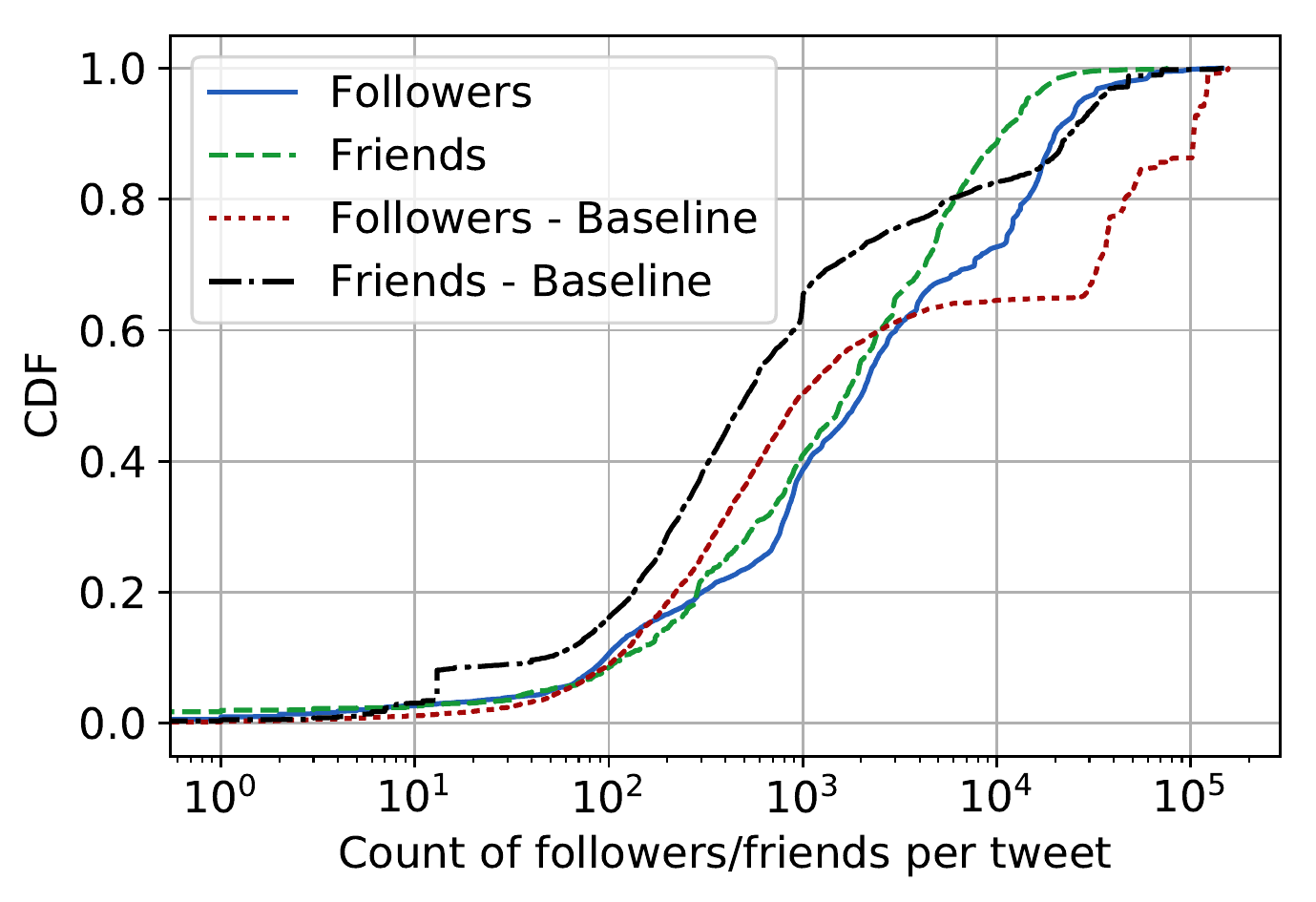}\label{subfig:cdf_followers}}
\subfigure[]{\includegraphics[width=0.49\columnwidth]{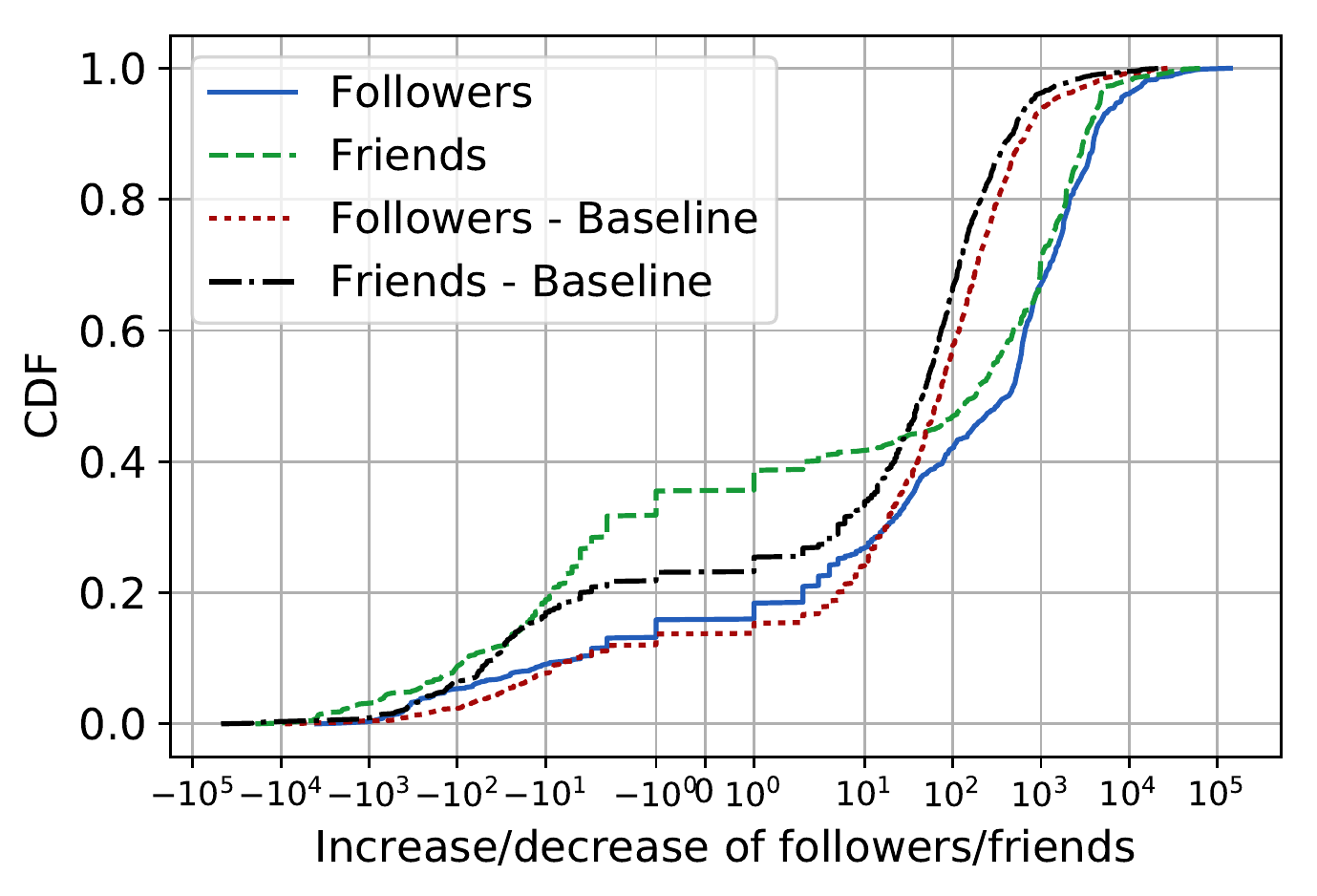}\label{subfig:cdf_followers_changes}}
\caption{CDF of the number of (a) followers/friends for each tweet and (b) increase in followers/friends for each user from the first to the last tweet.}
\label{fig:cdfs_domains_urls}
\end{figure}

\descr{Followers/Friends.} Next, we look at the number of followers and friends (i.e., the accounts one follows) of the Russian trolls, as this is an indication of the overall impact of a tweet.
In Fig.~\ref{subfig:cdf_followers}, we plot the CDF of the number of followers per tweet measured at the time of that tweet.
On average, Russian trolls have 7K followers and 3K friends, while our baseline has 25K followers and 6K friends.
We also note that in both samples, tweets reached a large number of Twitter users; at least 1K followers, with peaks up to 145K followers.
These results highlight that Russian trolls have a non-negligible number of followers, which can assist in pushing specific narratives to a much greater number of Twitter users.
We also assess the evolution of the Russian trolls in terms of the number of their followers and friends.
To this end, we get the follower and friend count for each user on their first and last tweet
and calculate the difference.
Fig.~\ref{subfig:cdf_followers_changes} plots the CDF of the increase/decrease of the followers and friends for each troll as well as random user in our baseline.
We observe that, on average, Russian trolls increase their number of followers and friends by 2,065 and 1,225, respectively, whereas for the baseline we observe an increase of 425 and 133 for followers and friends, respectively.
This suggests that Russian trolls work hard to increase their reachability within Twitter.

\begin{figure}[t]
\centering
\includegraphics[width=0.7\columnwidth]{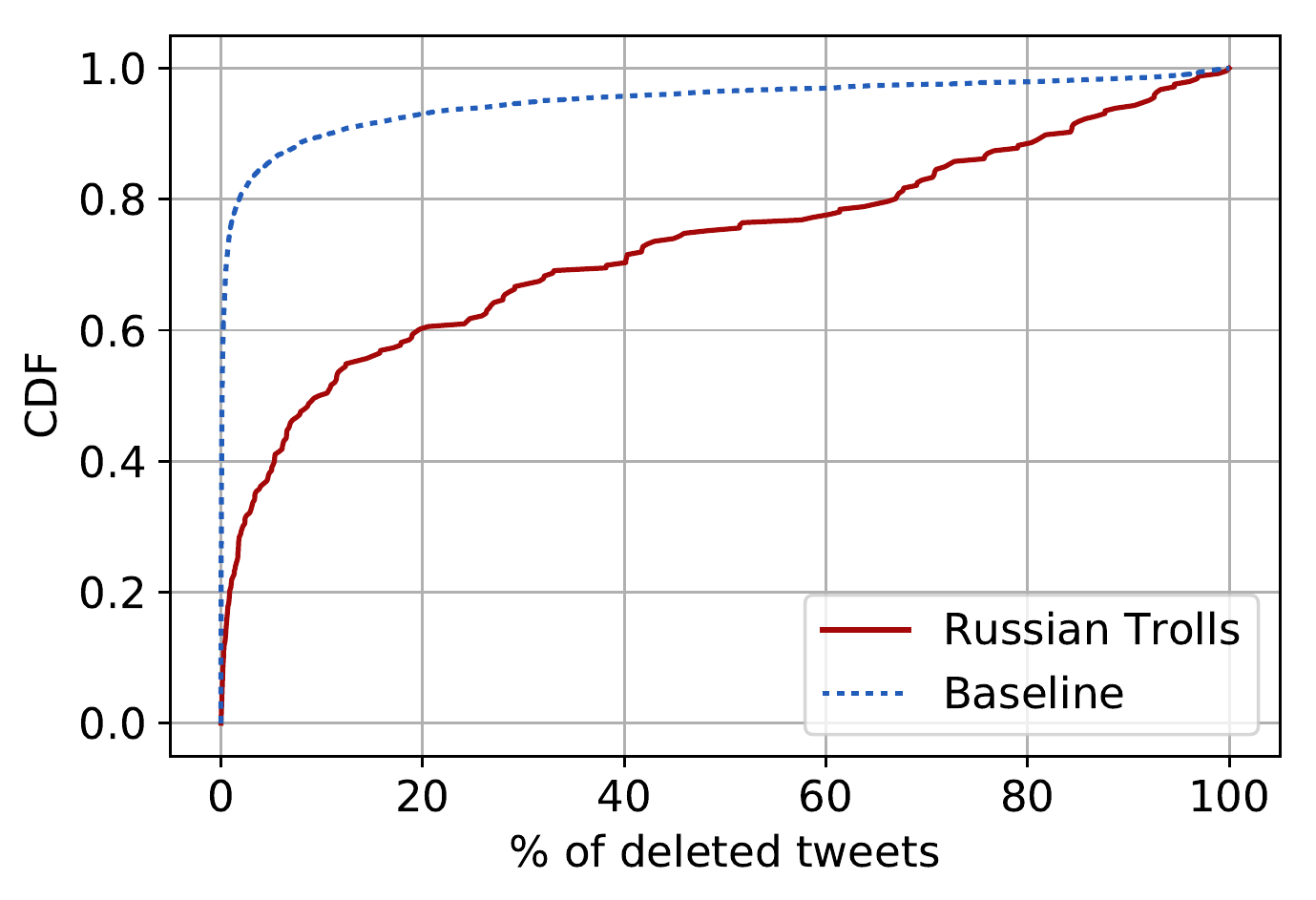}
\caption{CDF of the number of deleted tweets per observe deletion.}
\label{fig:cdf_deleted_tweets_user}
\end{figure}

\begin{figure}[t]
\centering
\includegraphics[width=0.9\columnwidth]{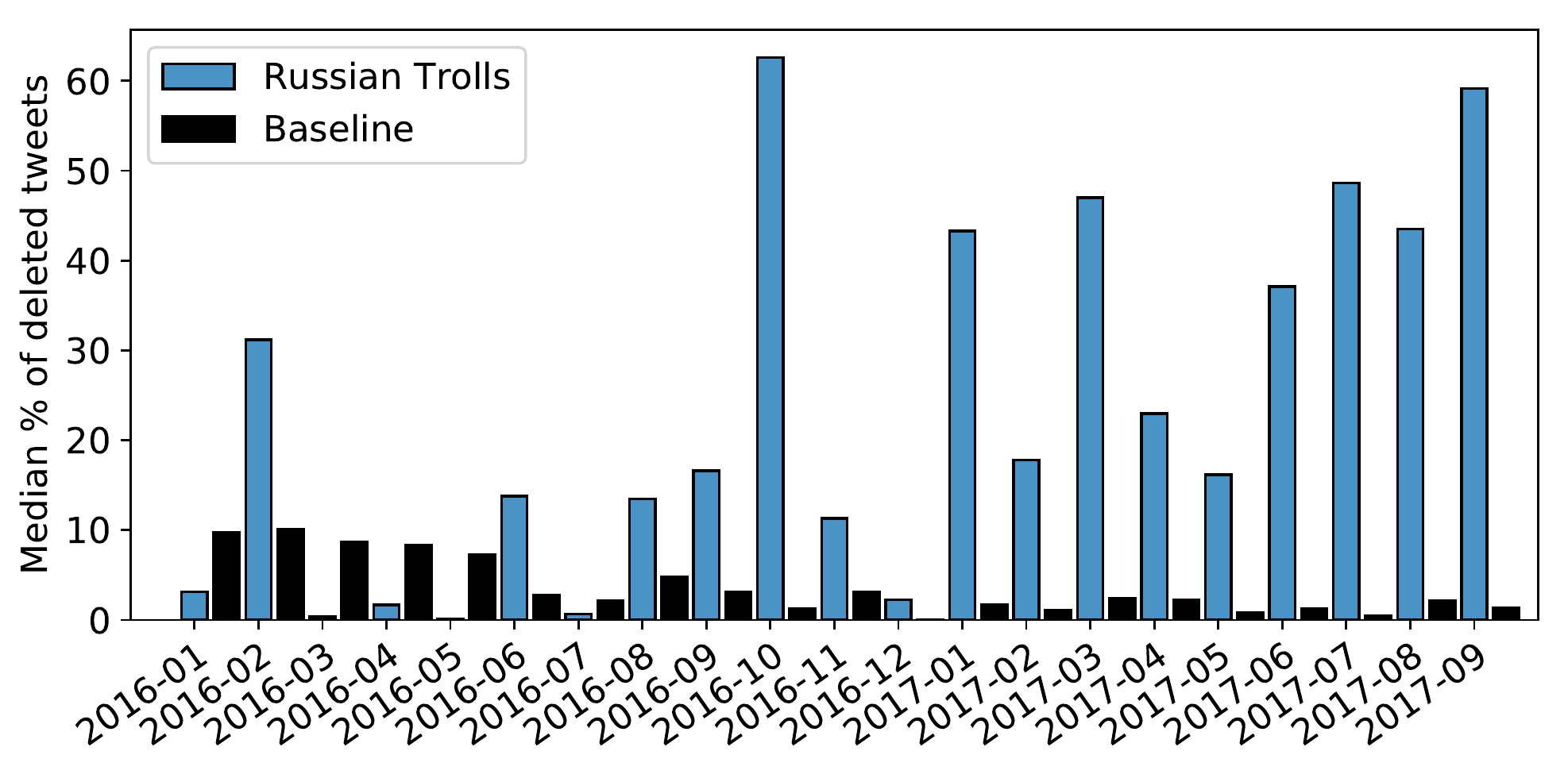}
  \caption{Average percentage of observed deletions per month. }
\label{fig:bc_deleted_tweets_month}
\end{figure}

\descr{Tweet Deletion.} Arguably, a reasonable strategy to avoid detection after posting tweets that aim to manipulate other users might be to delete them.
This is particularly useful when troll accounts change their identity and need to modify the narrative that they use to influence public opinion.
With each tweet, the Streaming API returns the total number of available tweets a user has up to that time.
Retrieving this count allows us to observe if a user has deleted a tweet, and around what period; we call this an ``observed deletion.''
Recall that our dataset is based on the 1\% sample of Twitter, thus, we can only estimate, in a conservative way, how many tweets are deleted; specifically, in between subsequent tweets, a user may have deleted and posted tweets that we do not observe. %
In Fig.~\ref{fig:cdf_deleted_tweets_user}, we plot the CDF of the number of deleted tweets per observed deletion. We observe that 13\% of the Russian trolls delete some of their tweets, with a median percentage of tweet deletion equal to 9.7\%.
Whereas, for the baseline set, 27\% of the accounts delete at least one tweet, but the median percentage is 0.1\%.
This means that the trolls delete their tweets in batches, possibly trying to cover their tracks or get a clean slate, while random users make a larger number of deletions but only a small percentage of their overall tweets, possibly because of typos.
We also report the distribution, over each month, of tweet deletions in Fig.~\ref{fig:bc_deleted_tweets_month}.
Specifically, we report the mean of the percentages for all observed deletions in our datasets.
Most of the tweets from Russian trolls are deleted in October 2016,
suggesting that these accounts attempted to get a clean slate a few months before the 2016 US elections.

\longVer{
\subsection{Case Study}
}
\shortVer{
\descr{Case Study.}} While the previous results provide a quantitative characterization of Russian trolls behavior, we believe there is value showing a concrete example of the behavior exhibited and how techniques played out.
We start on May 15, 2016, where the troll with screen name `Pen\_Air', was posing as a news account via its profile description: ``National American news.''
On September 8, 2016 as the US presidential elections approached, `Pen\_Air' became a Trump supporter, changing its screen name to `Blacks4DTrump' with a profile description of ``African-Americans stand with Trump to make America Great Again!''
Over the next 11 months, the account's tweet count grew from 49 to 642 while its follower count rose from 1.2K to \emph{nearly} 9K.
Then, around August 18, 2017, the account was seemingly repurposed.
Almost all of its previous tweets were deleted (the account's tweet count dropped to 35), it gained a new screen name (`southlonestar2'), and was now posing as a ``Proud American and TEXAN patriot! Stop ISLAM and PC. Don't mess with Texas'' according to its profile description.
When examining the accounts tweets, we see that most are clearly related to politics, featuring blunt right-wing attacks and ``talking points.''
For example, 
``Mr. Obama! Maybe you bring your girls and leave them in the bathroom with a grown man! \#bathroombill \#NObama $<$url$>$'' on May 15, 2016,
``\#HiLIARy has only two faces! And I hate both! \#NeverHillary \#Hillaryliesmatter $<$url$>$'' on May 19, 2016, and ``RT @TEN\_GOP: WikiLeaks \#DNCLeaks confirms something we all know: system is totally rigged! \#NeverHillary $<$url$>$.'' on July 22, 2016.

\longVer{
\subsection{Take-aways}
}
\shortVer{\descr{Take-aways.}} In summary, our analysis leads to the following observations.
First, we find evidence that trolls were actively involved in the dissemination of content related to world news and politics, as well as propaganda content regarding various topics such as ISIS and Islam. %
Moreover, several Russian trolls were created or repurposed a few weeks before notable world events, including the Republican National Convention meeting or the Charlottesville rally.
We also find that the trolls deleted a substantial amount of tweets in batches and overall made substantial changes to their accounts during the course of their lifespan.
Specifically, they changed their screen names aiming to pose as news outlets,
experienced significant rises in the numbers of followers and friends, etc.
Furthermore, our location analysis shows that Russian trolls might have tried to manipulate users  located in the USA, Germany, and possibly in their own country (i.e., Russia), by appearing to be located in those countries.
Finally, the fact that these accounts were active up until their recent suspension also highlights the need to develop more effective tools to detect such actors.

\begin{table}[t!]
\centering
\resizebox{0.9\columnwidth}{!}{%
\setlength{\tabcolsep}{0.4em} %
  \begin{tabular}{llrrrr}
  \toprule
      &        &     \textbf{/pol/} &  \textbf{Reddit} &   \textbf{Twitter} &  \textbf{Trolls} \\
  \midrule
\textbf{URLs}     & Russian state-sponsored &      6 &    13 &    19 &    19 \\
      & Other news sources &     47 &   168 &   159 &   192 \\
             &    All &    127 &   482 &   861 &   989 \\

  \midrule

\textbf{Events}   & Russian state-sponsored &     19 &    42 &      118 &    19 \\
      & Other news sources &    720 &  3,055 &     2,930 &   195 \\
           &     All &   1,685 &  9,531 &  1,537,612 &  1,461 \\

\midrule
\textbf{Mean $\lambda_0$}    &Russian state-sponsored &  0.0824 &  0.1865 &  0.2264 &  0.1228 \\
                & Other news sources &  0.0421 &  0.1447 &  0.1544 &  0.0663 \\
                        & All &  0.0324 &  0.1557 &  0.1553 &  0.0753 \\

\bottomrule
\end{tabular}
}
\caption{Total URLs with at least one event in Twitter, /pol/, Reddit, and Russian trolls on Twitter; total events for Russian state-sponsored news URLs, other news URLs and all the URLs; and mean background rate ($\lambda_0$) for each platform.}
\label{tbl:hawkes}
\end{table}

\section{Influence Estimation} \label{sec:influence}
Thus far, we have analyzed the behavior of the Russian trolls on the Twitter platform, and how this differs from that of a baseline of random users.
Allegedly, their main goal is to ultimately manipulate the opinion of other users and extend the cascade of disinformation they share (e.g., other users post similar content)~\cite{newsweek_manipulation}.
Therefore, we now set out to shed light on their impact, in terms of the dissemination of disinformation, on Twitter and on the greater Web.

To assess their influence, we look at the URLs posted by four groups of users: Russian trolls on Twitter, ``normal'' accounts on Twitter, Reddit users, and 4chan users (\dspol board).
For each unique URL, we fit a statistical model known as Hawkes Processes~\cite{linderman2014,lindermanArxiv}, which allows us to estimate the strength of connections between each of these four groups in terms of how likely an event -- the URL being posted by either trolls or normal users to a particular platform -- is to cause subsequent events in each of the groups.
For example, a strong connection from Reddit to \dspol would mean that a URL that appears on Reddit is likely to be seen and then re-posted on \dspol; whereas, a weak connection from trolls to normal users on Twitter indicates that a URL posted by trolls is less likely to be re-tweeted or re-posted by the latter.
We fit the Hawkes Processes using the methodology presented by~\cite{zannettou2017web}.

\begin{figure}[t]
\center
\hspace{-0.2cm}
\subfigure[All URLs]{\includegraphics[width=0.75\columnwidth]{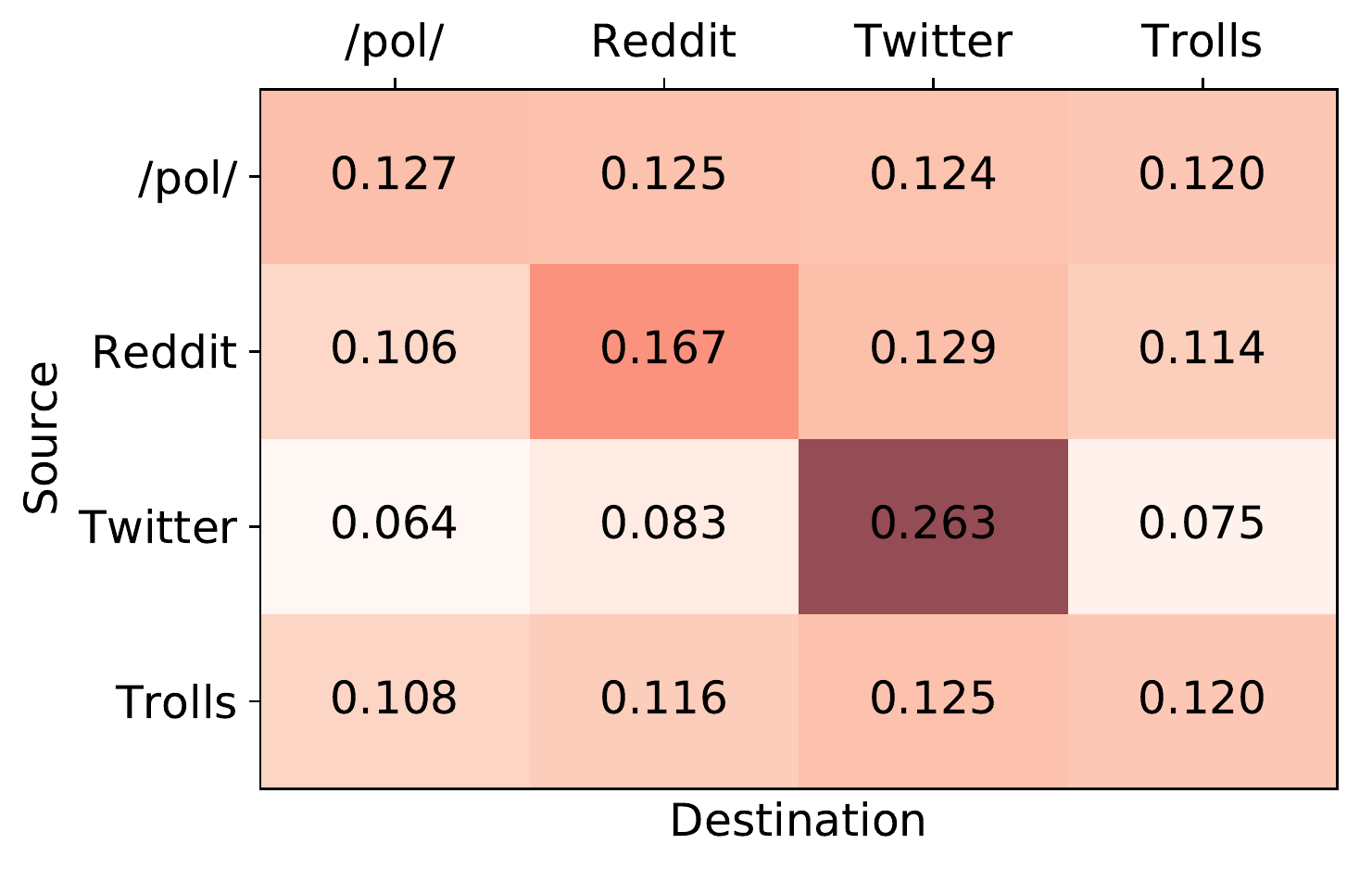}\label{subfig:weights_all}}
\hspace{-0.2cm}
\subfigure[News URLs]{\includegraphics[width=0.75\columnwidth]{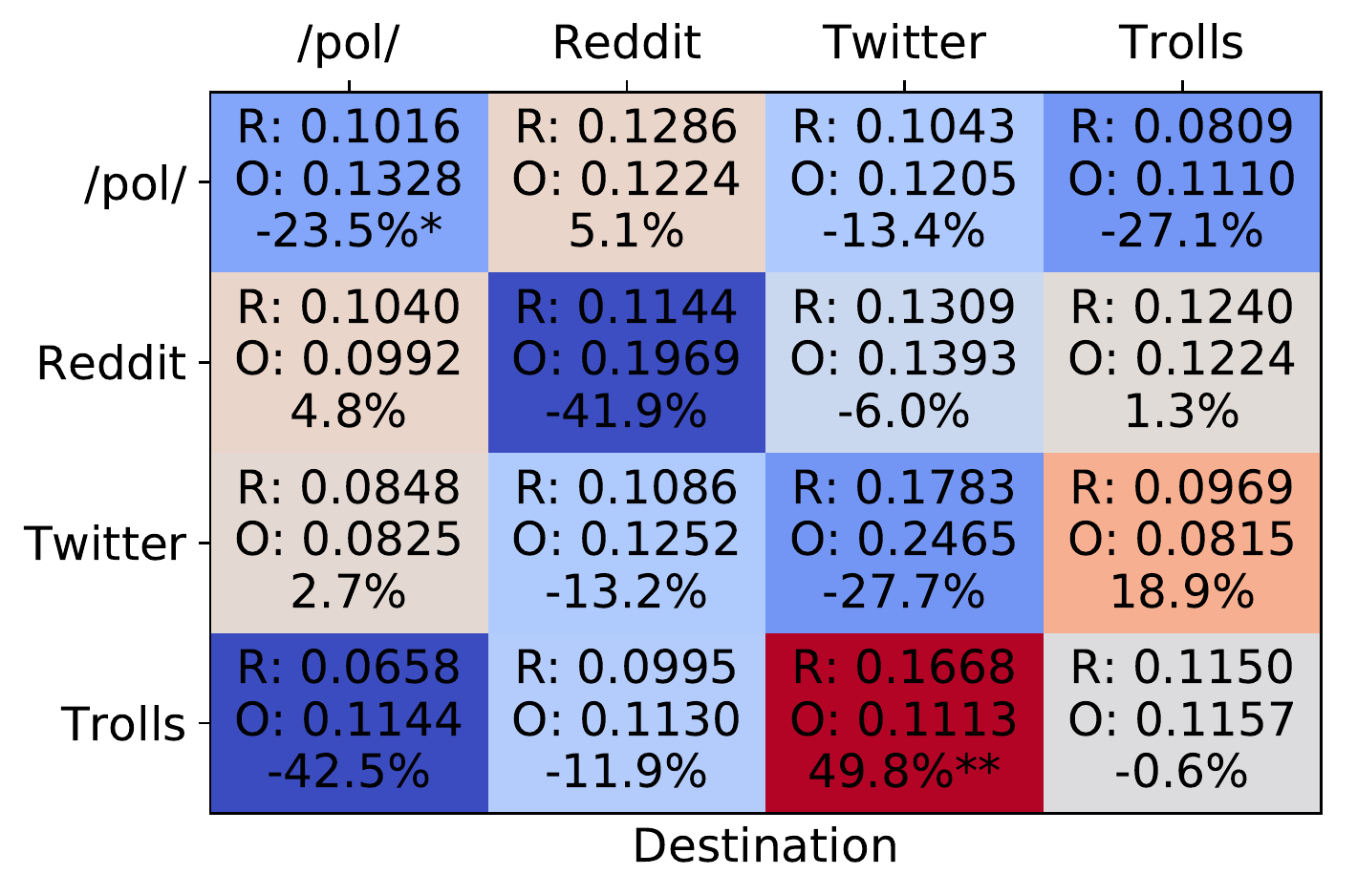}\label{subfig:weights_rt}}
\caption{Mean weights for (a) all URLs in our dataset and (b) news URLs categorized as Russian state-sponsored (R) and other mainstream and alternative news URLs (O). We also show the percent of increase/decrease between the two categories. Note that * and ** refer to statistical significance with, resp., $p< 0.05$ and $p < 0.01$.}
\label{fig:weights}
\end{figure}

To study the dissemination of different types of content, we look at three different sets of URLs: 1) The complete set of all URLs posted by Russian trolls; 2) The subset of URLs for Russian state-sponsored news, namely, RT (Russia Today); and 3) The subset of URLs from other mainstream and alternative news sources using the list provided by~\cite{zannettou2017web}.
Table~\ref{tbl:hawkes} summarizes the number of URLs, number of events (i.e., occurrences of a given URL) as well as the mean background rate for each category and social network.
The background rate defines the rate at which events occur excluding the influence of the platforms included in the model; the background rate includes events created spontaneously on each platform, such as by a user sharing the article from the original source, or those generated by another platform not monitored by us like Facebook.
The number of events for Russian state-sponsored news sources is substantially lower than the number of events from other news sources.
This is expected since the former only includes one news source (RT),
however, it is interesting that the background rates for these URLs are higher than for other news sources, meaning that events from Russian state-sponsored news are more likely to occur spontaneously.%

Fitting a Hawkes model yields a weight matrix, which characterizes the strength of the connections between the groups we study.
Each weight value represents the connection strength from one group to another and can be interpreted as the expected number of subsequent events that will occur on the second group after each event on the first.
The mean weight values over all URLs, as well as for the URLs from RT and other news URLs, are presented in Fig.~\ref{fig:weights}.
We observe that for \dspol, Reddit, and normal users on Twitter, the greatest weights are from each group to itself, meaning that reposts/retweets on the same site are more common than sharing the URL to the other platforms (Fig.~\ref{subfig:weights_all}).
For the Russian Trolls on Twitter, however, the weight is greater from the trolls to Twitter than from the trolls to themselves, perhaps reflecting their use as an avenue for \textit{disseminating} information to normal Twitter users (Fig.~\ref{subfig:weights_rt}).
Also, we observe that, in most cases, the connections are stronger for non-Russia state-sponsored news, indicating that regular users are more inclined to share news articles from mainstream and alternative news sources.

\begin{figure}[t]
\center
\hspace{-0.2cm}
\subfigure[All URLs]{\includegraphics[width=0.75\columnwidth]{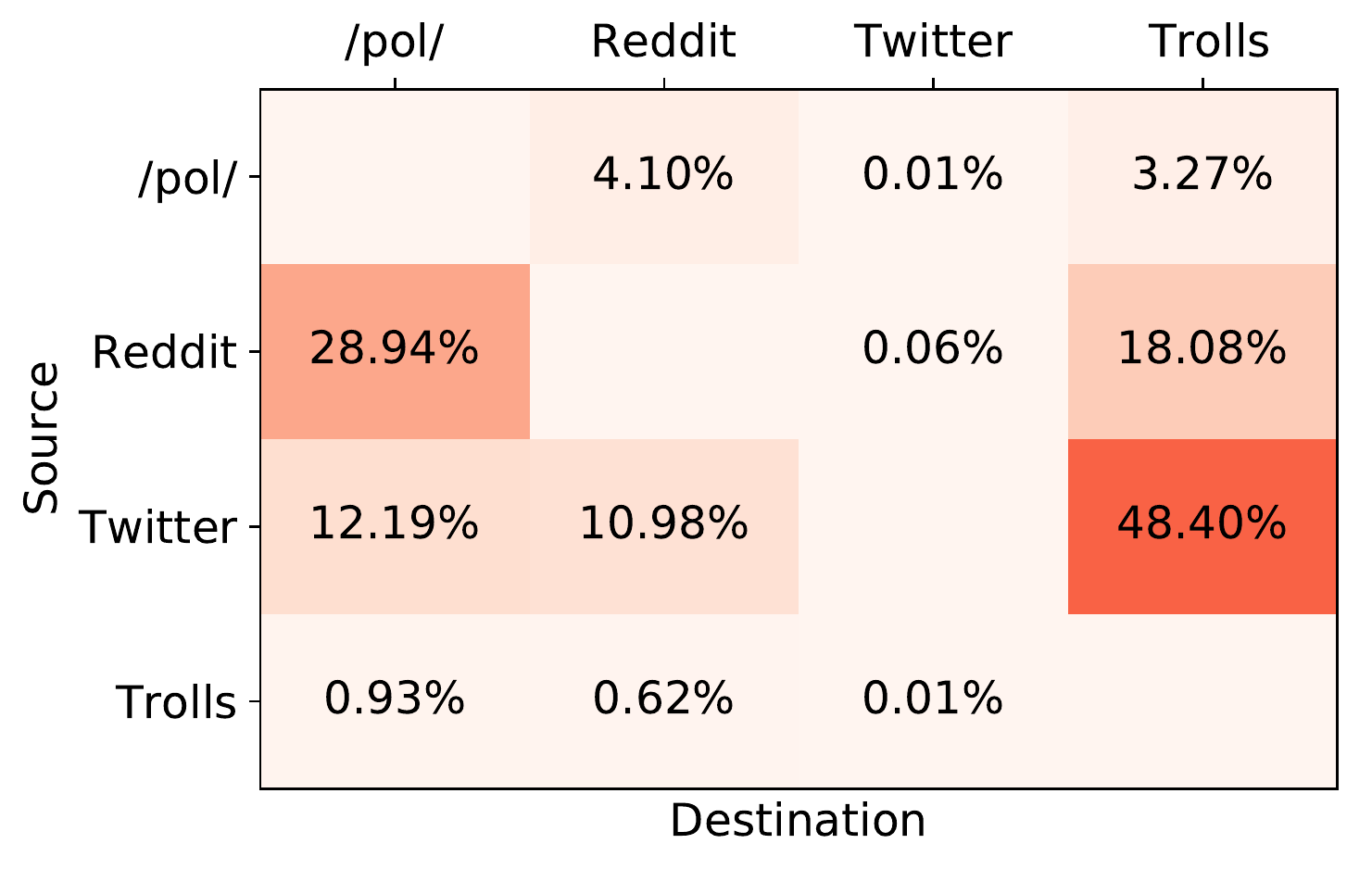}\label{subfig:expected_all}}
\hspace{-0.2cm}
\subfigure[News URLs]{\includegraphics[width=0.75\columnwidth]{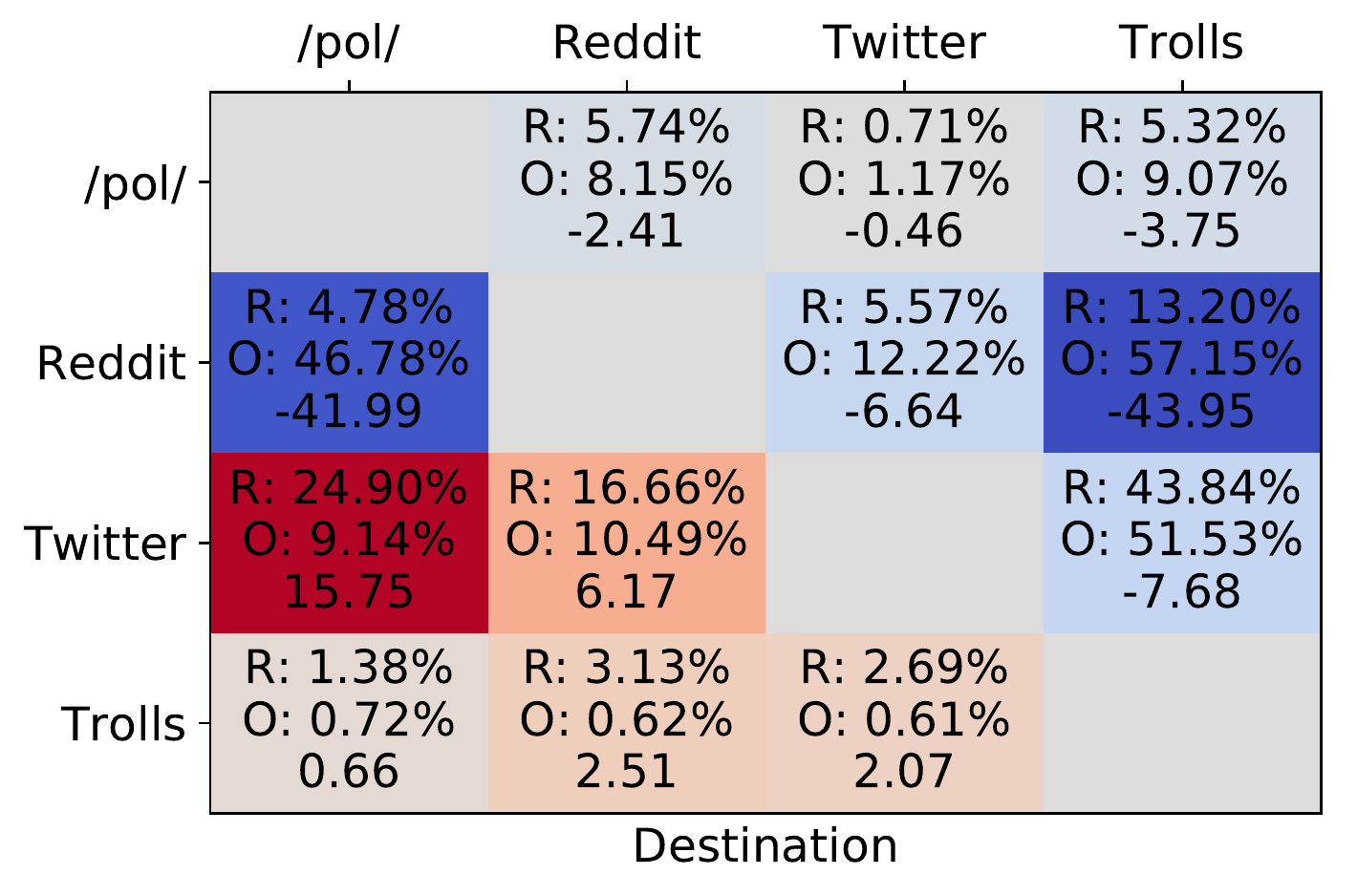}\label{subfig:expected_rt}}
\caption{Estimated mean percentages of events created because of other events for (a) all URLs  and (b) Russian state-sponsored URLs (R) and other mainstream and alternative news URLs (O). We also show the difference between the two categories of news.}
\label{fig:expected_percentages}
\end{figure}

Looking at the Russian trolls and normal Twitter users, we see that the trolls are more likely to retweet Russian state-sponsored URLs from normal Twitter users than other news sources;  conversely, normal Twitter users are more likely to retweet Russian state-sponsored URLs from the troll accounts.
In order to assess the significance of our results, we perform two-sample Kolmogorov-Smirnov tests on the weight distributions for the RT URLs and the other news URLs for each source-destination platform pair (depicted as stars in the Fig.~\ref{subfig:weights_rt}). Small $p$ value means there is a statistically significant difference in the way that RT URLs propagate from the source to the destination platform.
Most of the source-destination pairs have no statistical significance, however for the Russian trolls--Twitter users pair, we find \longVer{statistically} significance difference with $p < 0.01$.

In Fig.~\ref{fig:expected_percentages}, we report the estimated total impact for each pair of platforms, for both Russian state-sponsored news, other news sources as well as all the observed URLs.
We determine the impact by calculating, based on the estimated weights and the number of events, the percentage of events on a destination platform caused by events on a source platform,
following the same methodology as~\cite{zannettou2017web}.
For all URLs (Fig.~\ref{subfig:expected_all}), we find that the influence of Russian trolls is negligible on Twitter (0.01\%), while for \dspol and Reddit it is slightly higher (0.93\% and 0.62\%, respectively). For other pairs, the larger impacts are between Reddit--\dspol and Twitter-Russian trolls, mainly due to the larger population of users.
Looking at the estimated impact for RT and other news sources (Fig.~\ref{subfig:expected_rt}), we note that the trolls influenced the other platforms approximately the same for alternative and mainstream news sources (0.72\%, 0.62\%, and 0.61 for \dspol, Reddit, and Twitter, respectively).
Interestingly, Russian trolls have a much larger impact on all the other platforms for the RT news when compared to the other news sources: approximately 2 times more on \dspol, 5 times more on Reddit, and 4 times more on Twitter.

\descr{Take-aways.}
Using Hawkes processes, we were able to assess the degree of influence Russian trolls had on Twitter, Reddit, and \dspol by examining the diffusion of information via URLs to news.
Our results indicate that their influence is actually quite limited.
With the exception of news originating from the Russian state-sponsored news outlet RT, the troll accounts were generally less influential than other users on Reddit, Twitter, and 4chan.
However, our analysis is based only on 1K troll accounts found in Twitter's 1\% stream, and, as mentioned previously, Twitter recently announced they had discovered over 1K more trolls and more than 50K automated accounts.
With that in mind, there are several potential explanations behind this limited influence.
For example, there might be a lot of influence attributed to regular Twitter users that belongs to newly announced troll accounts.
Considering that Twitter announced the discovery of ``only'' 1K more troll accounts, we suspect that this is not really the case.
Another, more plausible explanation is that the troll accounts are just not terribly efficient at spreading news, and instead are more concerned with causing havoc by pushing ideas, engaging other users, or even taking both sides of controversial online discussions~\cite{steward2018examining}.
This scenario makes more sense considering that, along with 1K new troll accounts, Twitter also announced discovering over 50K \emph{automated} accounts that might be more efficient in terms of driving traffic to specific URLs.
\section{Related Work} \label{sec:related}
We now review previous work on opinion manipulation, politically motivated disinformation on the Web, and analysis of state-sponsored trolls' behavior. %

\descr{Opinion manipulation.}
 The practice of swaying opinion on the Web is a long-standing issue as malicious actors attempt to push their agenda.
Kumar et al.~\cite{kumar2017sockpuppets} study how users create multiple accounts, called \emph{sockpuppets}, that participate in Web communities to manipulate users' opinions. They find that sockpuppets exhibit different posting behavior when compared to benign users.
Mihaylov et al.~\cite{mihaylov2015finding} show that trolls can manipulate users' opinions in forums, while in their follow-up work~\cite{mihaylov2016hunting}  they highlight the two types of manipulation trolls: those paid to operate and those that are called out as such by other users.
Then, Volkova and Bell~\cite{volkova2016account} predict the deletion of Twitter accounts because they are trolls,
focusing on those that shared content related to the Russian-Ukraine crisis. %
Elyashar et al.~\cite{elyashar2017is} distinguish authentic discussions from campaigns to manipulate the public's opinion,
using a set of similarity functions alongside historical data. %
Finally, Varol et al.~\cite{varol2017early} identify memes %
 that become popular due to \emph{coordinated} efforts,
and achieve %
0.75 AUC score before memes become trending and 0.95 afterwards.

\descr{False information on the political stage.}
Conover et al.~\cite{conover2011political} study the interactions of right- and left-leaning communities on Twitter during the 2010 US midterm elections,
finding that the retweet network has limited connectivity between the two communities, which does not happen in the mentions network;
mainly because users engage others users with different ideologies and expose them to different opinions. %
Ratkiewicz et al.~\cite{ratkiewicz2011detecting} use machine learning to detect the early stages of false political information spreading on Twitter and introduce a framework that considers topological, content-based, and crowdsourced features of the information diffusion.
Wong et al.~\cite{wong2013quantifying} quantify political leaning of users and news outlets during the 2012 US presidential election on Twitter by using an inference engine that considers tweeting and retweeting behavior of articles.
Yang et al.~\cite{yang2016social} investigate the topics of discussions on Twitter for 51 US political persons
showing that Democrats and Republicans are active in a similar way on Twitter.
Le et al.~\cite{le2017revisiting} study 50M tweets regarding the 2016 US election primaries and highlight the importance of three factors in political discussions on social media, namely the \textit{party}, 
\textit{policy considerations}, 
and \textit{personality} of the candidates.
Howard and Kollanyi~\cite{howard2016bots} study the role of bots in Twitter conversations during  the 2016 Brexit referendum.  By analyzing 1.5M tweets, they find that most tweets are in favor of Brexit and that there are bots with various levels of automation.%
Also, Hegelich and Janetzko~\cite{hegelich2016are} highlight that bots have a political agenda and that they exhibit various behaviors, e.g., trying to hide their identity and promoting topics through hashtags and retweets.
Finally, a large body of work focuses on social bots~\cite{bessi2016social,davis2016bot,ferrara2016the,ferrara2017disinformation,varol2017online} and their role in spreading disinformation, highlighting that they can manipulate the public's opinion at large scale, thus potentially affecting the outcome of political elections.

\descr{State-sponsored trolls.}
Recent work aim to uncover the behavior of state-sponsored trolls on the Web by analyzing ground truth datasets identified independently by social network operators like Facebook and Twitter.
Specifically, Zannettou et al.~\cite{zannettou2018let} analyze a set of 10M posts by Russian and Iranian trolls on Twitter shedding light into their campaigns, their targets, and how their campaigns/targets/behavior change over time.
In follow-up work, Zannettou et al.~\cite{zannettou2019characterizing} focus on the dissemination of images by Russian trolls on Twitter.
They analyze 1.8M images finding that the dissemination of images is tightly coupled with real-world events and that the shared images were mainly related to politics and world news with a focus on Russia and USA.
Also, Steward et al.~\cite{steward2018examining} focus on discussions related to the Black Lives Matter movement and how content from Russian trolls was retweeted by other users.
Using the retweet network, they find the existence of two communities; one left- and one right-leaning communities. Also, they note that trolls infiltrated both communities, setting out to push specific narratives.
Im et al.~\cite{im2019still} focus on detecting Twitter users that likely act on behalf of the Russian Internet Research Agency (i.e., they are Russian trolls). 
To do this, they use conventional machine learning techniques with a set of handcrafted features. 
By running the proposed classifier to an out-of-sample set of users they find that Russian trolls are still very active on the Web.
Kim et al~\cite{kim2019tracking} study the temporal traces of Russian trolls on Twitter finding cooperation between several trolls, substantial interplay between right and left leaning trolls, and that trolls have multiple agendas to disturb democracy in western countries.
Finally, Boyd et al.~\cite{boyd2018characterizing} perform a linguistic analysis on tweets from Russian trolls during the 2016 US elections: they find that trolls were targeting differently right and left leaning communities on Twitter and that their tweets were linguistically unique when compared to controlled-sample of english-speaking accounts.

\section{Conclusion} \label{sec:conclusion}
In this paper, we analyzed the behavior and use of the Twitter platform by Russian trolls during the course of 21 months.
We showed that Russian trolls exhibited interesting differences when compared with a set of random users, actively disseminated politics-related content, adopted multiple identities during their account's lifespan, and that they aimed to increase their impact on Twitter by increasing their followers.
Also, we quantified the influence that Russian trolls have on Twitter, Reddit, and \dspol using a statistical model known as Hawkes Processes.
Our findings show that trolls' influence was not substantial with respect to the other platforms, with the significant exception of news published by the Russian state-sponsored news outlet RT.

\descr{Acknowledgments.}
This project has received funding from the European Union's Horizon 2020 Research and Innovation program under the Marie Sk\l{}odowska Curie ENCASE project (GA No. 691025) and under the Cybersecurity CONCORDIA project (GA No. 830927).

\small
\bibliographystyle{abbrv}

\begin{thebibliography}{10}

\bibitem{bessi2016social}
A.~Bessi and E.~Ferrara.
\newblock {Social bots distort the 2016 US Presidential election online
  discussion}.
\newblock {\em First Monday}, 21(11), 2016.

\bibitem{boyd2018characterizing}
R.~L. Boyd, A.~Spangher, A.~Fourney, B.~Nushi, G.~Ranade, J.~W. Pennebaker, and
  E.~Horvitz.
\newblock Characterizing the internet research agency’s social media
  operations during the 2016 us presidential election using linguistic
  analyses.
\newblock {\em PsyArXiv. October}, 1, 2018.

\bibitem{conover2011political}
M.~Conover, J.~Ratkiewicz, M.~R. Francisco, B.~Goncalves, F.~Menczer, and
  A.~Flammini.
\newblock {Political Polarization on Twitter}.
\newblock In {\em ICWSM}, 2011.

\bibitem{davis2016bot}
C.~A. Davis, O.~Varol, E.~Ferrara, A.~Flammini, and F.~Menczer.
\newblock {BotOrNot: A System to Evaluate Social Bots}.
\newblock In {\em WWW}, 2016.

\bibitem{newsweek_manipulation}
S.~Earle.
\newblock {Trolls, Bots and Fake News: The Mysterious World of Social Media
  Manipulation}.
\newblock \url{https://goo.gl/nz7E8r}, 2017.

\bibitem{egele2017towards}
M.~Egele, G.~Stringhini, C.~Kruegel, and G.~Vigna.
\newblock {Towards detecting compromised accounts on social networks}.
\newblock {\em IEEE TDSC}, 2017.

\bibitem{elyashar2017is}
A.~Elyashar, J.~Bendahan, and R.~Puzis.
\newblock {Is the Online Discussion Manipulated? Quantifying the Online
  Discussion Authenticity within Online Social Media}.
\newblock {\em CoRR}, abs/1708.02763, 2017.

\bibitem{ferrara2017disinformation}
E.~Ferrara.
\newblock {Disinformation and social bot operations in the run up to the 2017
  French presidential election}.
\newblock {\em ArXiv 1707.00086}, 2017.

\bibitem{ferrara2016the}
E.~Ferrara, O.~Varol, C.~A. Davis, F.~Menczer, and A.~Flammini.
\newblock {The rise of social bots}.
\newblock {\em Commun. ACM}, 2016.

\bibitem{gupta2012credibility}
A.~Gupta and P.~Kumaraguru.
\newblock {Credibility ranking of tweets during high impact events}.
\newblock In {\em PSOSM '12}, 2012.

\bibitem{hegelich2016are}
S.~Hegelich and D.~Janetzko.
\newblock {Are Social Bots on Twitter Political Actors? Empirical Evidence from
  a Ukrainian Social Botnet}.
\newblock In {\em ICWSM}, 2016.

\bibitem{hine2016longitudinal}
G.~E. Hine, J.~Onaolapo, E.~De~Cristofaro, N.~Kourtellis, I.~Leontiadis,
  R.~Samaras, G.~Stringhini, and J.~Blackburn.
\newblock {Kek, Cucks, and God Emperor Trump: A Measurement Study of 4chan's
  Politically Incorrect Forum and Its Effects on the Web}.
\newblock In {\em ICWSM}, 2017.

\bibitem{howard2016bots}
P.~N. Howard and B.~Kollanyi.
\newblock {Bots, \#StrongerIn, and \#Brexit: Computational Propaganda during
  the UK-EU Referendum}.
\newblock {\em Arxiv 1606.06356}, 2016.

\bibitem{im2019still}
J.~Im, E.~Chandrasekharan, J.~Sargent, P.~Lighthammer, T.~Denby, A.~Bhargava,
  L.~Hemphill, D.~Jurgens, and E.~Gilbert.
\newblock {Still out there: Modeling and Identifying Russian Troll Accounts on
  Twitter}.
\newblock {\em arXiv preprint 1901.11162}, 2019.

\bibitem{kim2019tracking}
D.~Kim, T.~Graham, Z.~Wan, and M.-A. Rizoiu.
\newblock {Tracking the Digital Traces of Russian Trolls: Distinguishing the
  Roles and Strategy of Trolls On Twitter}.
\newblock {\em arXiv preprint 1901.05228}, 2019.

\bibitem{kumar2017sockpuppets}
S.~Kumar, J.~Cheng, J.~Leskovec, and V.~S. Subrahmanian.
\newblock {An Army of Me: Sockpuppets in Online Discussion Communities}.
\newblock In {\em WWW}, 2017.

\bibitem{le2017revisiting}
H.~T. Le, G.~R. Boynton, Y.~Mejova, Z.~Shafiq, and P.~Srinivasan.
\newblock {Revisiting The American Voter on Twitter}.
\newblock In {\em CHI}, 2017.

\bibitem{linderman2014}
S.~W. Linderman and R.~P. Adams.
\newblock {Discovering Latent Network Structure in Point Process Data}.
\newblock In {\em ICML}, 2014.

\bibitem{lindermanArxiv}
S.~W. Linderman and R.~P. Adams.
\newblock {Scalable Bayesian Inference for Excitatory Point Process Networks}.
\newblock {\em ArXiv 1507.03228}, 2015.

\bibitem{mariconti2017s}
E.~Mariconti, J.~Onaolapo, S.~S. Ahmad, N.~Nikiforou, M.~Egele, N.~Nikiforakis,
  and G.~Stringhini.
\newblock {What's in a Name?: Understanding Profile Name Reuse on Twitter}.
\newblock In {\em WWW}, 2017.

\bibitem{mihaylov2015finding}
T.~Mihaylov, G.~Georgiev, and P.~Nakov.
\newblock {Finding Opinion Manipulation Trolls in News Community Forums}.
\newblock In {\em CoNLL}, 2015.

\bibitem{mihaylov2016hunting}
T.~Mihaylov and P.~Nakov.
\newblock {Hunting for Troll Comments in News Community Forums}.
\newblock In {\em ACL}, 2016.

\bibitem{ratkiewicz2011detecting}
J.~Ratkiewicz, M.~Conover, M.~R. Meiss, B.~Gonçalves, A.~Flammini, and
  F.~Menczer.
\newblock {Detecting and Tracking Political Abuse in Social Media}.
\newblock In {\em ICWSM}, 2011.

\bibitem{smedt2012pattern}
T.~D. Smedt and W.~Daelemans.
\newblock {Pattern for python}.
\newblock {\em Journal of Machine Learning Research}, 2012.

\bibitem{starbird2017examining}
K.~Starbird.
\newblock {Examining the Alternative Media Ecosystem Through the Production of
  Alternative Narratives of Mass Shooting Events on Twitter}.
\newblock In {\em ICWSM}, 2017.

\bibitem{steward2018examining}
L.~Steward, A.~Arif, and K.~Starbird.
\newblock {Examining Trolls and Polarization with a Retweet Network}.
\newblock In {\em MIS2}, 2018.

\bibitem{independent}
{The Independent}.
\newblock {St Petersburg 'troll farm' had 90 dedicated staff working to
  influence US election campaign}.
\newblock \url{https://ind.pn/2yuCQdy}, 2017.

\bibitem{varol2017online}
O.~Varol, E.~Ferrara, C.~A. Davis, F.~Menczer, and A.~Flammini.
\newblock {Online Human-Bot Interactions: Detection, Estimation, and
  Characterization}.
\newblock In {\em ICWSM}, 2017.

\bibitem{varol2017early}
O.~Varol, E.~Ferrara, F.~Menczer, and A.~Flammini.
\newblock {Early detection of promoted campaigns on social media}.
\newblock {\em EPJ Data Science}, 2017.

\bibitem{volkova2016account}
S.~Volkova and E.~Bell.
\newblock {Account Deletion Prediction on RuNet: A Case Study of Suspicious
  Twitter Accounts Active During the Russian-Ukrainian Crisis}.
\newblock In {\em NAACL-HLT}, 2016.

\bibitem{wong2013quantifying}
F.~M.~F. Wong, C.-W. Tan, S.~Sen, and M.~Chiang.
\newblock {Quantifying Political Leaning from Tweets and Retweets}.
\newblock In {\em ICWSM}, 2013.

\bibitem{yang2016social}
X.~Yang, B.-C. Chen, M.~Maity, and E.~Ferrara.
\newblock {Social Politics: Agenda Setting and Political Communication on
  Social Media}.
\newblock In {\em SocInfo}, 2016.

\bibitem{zannettou2019characterizing}
S.~Zannettou, B.~Bradlyn, E.~De~Cristofaro, G.~Stringhini, and J.~Blackburn.
\newblock {Characterizing the Use of Images by State-Sponsored Troll Accounts
  on Twitter}.
\newblock {\em arXiv preprint 1901.05997}, 2019.

\bibitem{zannettou2017web}
S.~Zannettou, T.~Caulfield, E.~De~Cristofaro, N.~Kourtellis, I.~Leontiadis,
  M.~Sirivianos, G.~Stringhini, and J.~Blackburn.
\newblock {The Web Centipede: Understanding How Web Communities Influence Each
  Other Through the Lens of Mainstream and Alternative News Sources}.
\newblock In {\em ACM IMC}, 2017.

\bibitem{zannettou2018let}
S.~Zannettou, T.~Caulfield, W.~Setzer, M.~Sirivianos, G.~Stringhini, and
  J.~Blackburn.
\newblock {Who Let the Trolls out? Towards Understanding State-Sponsored
  Trolls}.
\newblock {\em arXiv preprint 1811.03130}, 2018.

\end{thebibliography}

\end{document}